\documentclass[a4paper,fleqn,usenatbib]{mnras}
\usepackage{enumerate}
\usepackage[T1]{fontenc}
\usepackage{ae,aecompl}
\usepackage{indentfirst}
\usepackage{graphicx}
\usepackage{amsmath}
\usepackage{amssymb}
\usepackage{color}
\usepackage{mathptmx}
\usepackage{epsfig}
\usepackage{wasysym}
\usepackage{xfrac}
\usepackage{pifont}
\usepackage[table,xcdraw]{xcolor}
\usepackage{booktabs}
\usepackage{multirow}
\usepackage{comment}
\usepackage{longtable}
\usepackage[normalem]{ulem}
\usepackage{subcaption}
\title[Dwarf galaxy morphologies]{The morphological mix of dwarf galaxies in the nearby Universe}

\author[I. Lazar et al.]{
I. Lazar$^{1}$\thanks{E-mail: i.lazar@herts.ac.uk}, S. Kaviraj$^{1}$, A. E. Watkins$^{1}$, G. Martin$^{2,3,4}$, B. Bichang'a$^{1}$ and R. A. Jackson$^{5}$\\
$^{1}$Centre for Astrophysics Research, University of Hertfordshire, College Lane, Hatfield AL10 9AB, UK\\
$^{2}$School of Physics and Astronomy, University of Nottingham, University Park, Nottingham NG7 2RD, UK\\
$^{3}$Korea Astronomy and Space Science Institute, 776 Daedeokdae-ro, Yuseong-gu, Daejeon 34055, Korea\\
$^{4}$Steward Observatory, University of Arizona, 933 N. Cherry Ave, Tucson, AZ 85719, USA\\
$^{5}$Department of Physics and Astronomy, University of Victoria, Victoria, BC, Canada V8P 5C2\\
}
\pubyear{2024}

\begin{document}
\label{firstpage}
\pagerange{\pageref{firstpage}--\pageref{lastpage}}
\maketitle

\begin{abstract}
We use a complete, unbiased sample of 257 dwarf (10$^{8}$ M$_{\odot}$ < \textit{M}$_{\rm{\star}}$ < 10$^{9.5}$ M$_{\odot}$) galaxies at $z<0.08$, in the COSMOS field, to study the morphological mix of the dwarf population {\color{black}in low-density environments}. Visual inspection of extremely deep optical images and their unsharp-masked counterparts reveals three principal dwarf morphological classes. 43 and 45 per cent of dwarfs exhibit the traditional `early-type' (elliptical/S0) and `late-type' (spiral) morphologies respectively. However, 10 per cent populate a `featureless' class, that lacks both the central light concentration seen in early-types and any spiral structure - this class is missing in the massive-galaxy regime. 14, 27 and 19 per cent of early-type, late-type and featureless dwarfs respectively show evidence for interactions, which drive around 20 per cent of the overall star formation activity in the dwarf population. Compared to their massive counterparts, dwarf early-types show a much lower incidence of interactions, are significantly less concentrated and share similar rest-frame colours as dwarf late-types. This suggests that the formation histories of dwarf and massive early-types are different, with dwarf early-types being shaped less by interactions and more by secular processes. The lack of large groups or clusters in COSMOS at $z<0.08$, and the fact that our dwarf morphological classes show similar local density, suggests that featureless dwarfs in low-density environments are created via internal baryonic feedback, rather than by environmental processes. Finally, while interacting dwarfs can be identified using the asymmetry parameter, it is challenging to cleanly separate early and late-type dwarfs using traditional morphological parameters, such as `CAS', \textit{M}$_{20}$ and the Gini coefficient (unlike in the massive-galaxy regime).
\end{abstract}

\begin{keywords}
galaxies: formation -- galaxies: evolution -- galaxies: dwarf -- galaxies: structure 
\end{keywords}

%--------------------------------------------------------------

\section{Introduction}

Morphology is a fundamental parameter in observational astrophysics, which is strongly correlated with the evolution of galaxies. In the massive-galaxy (\textit{M}$_\star$ > 10$^{9.5}$ M$_\odot$) regime, the principal morphological classes are well described by the classical `Hubble sequence' \citep{Hubble1936}: `early-type' galaxies (ETGs), which include S0 and elliptical systems, and `late-type' galaxies (LTGs), which include different types of disc (spiral) galaxies. Many studies have shown that, in massive galaxies, these morphological types differ not only in their structural characteristics but also in their physical properties, such as colour, star formation rate (SFR) and stellar mass \citep[e.g.][]{Conselice2014,Martin2018_sph,Sampaio2022}. ETGs, which are more pressure-supported than LTGs, typically exhibit low SFRs, red colours and high stellar masses \citep[e.g.][]{Bernardi2003,Kaviraj2007,Cappellari2013}. LTGs, which are largely supported by rotation (and exhibit spiral arms as a result), typically exhibit higher SFRs, blue colours and lower stellar masses. 

Morphology encodes key information about the dominant physical processes that have shaped the formation of galaxies. The general consensus, in the massive-galaxy regime, is that LTGs largely form via secular cold gas accretion \citep{White1978,Fall1980} and experience either a relatively quiet history \citep[e.g.][]{Jackson2022} or a larger fraction of gas-rich (and mainly prograde) mergers that helps maintain their discy morphologies and relatively high SFRs over their lifetimes \citep[e.g.][]{Martin2018_sph}. ETGs, on the other hand, are thought to be the end products of multiple minor or major mergers \citep[e.g.][]{Martin2018_sph} and, in high density environments, also experience processes that promote gas depletion, such as tidal interactions \citep{Moore1998,Jackson2021a} and ram pressure stripping \citep{Walker1996,Martin2019,Jackson2021a}.    

A wide array of methods have traditionally been used to determine the morphology of galaxies \citep[where some notable innovative studies include the work done in e.g.][]{Vaucouleurs1959,Sandage1961,Buta1996}. Direct visual inspection of galaxy images, whilst the most time-consuming, produces accurate classifications \citep[e.g.][]{Lintott2011,Kaviraj2014}, against which automated methods are often calibrated. The classification of galaxies in large surveys typically requires automation, which is often achieved using both parametric and non-parametric methods. A commonly used parametric method for distinguishing between early and late-type galaxies is fitting S\'ersic indices \citep{Sersic1963} using galaxy surface brightness profiles \citep[e.g.][]{Bottrell2019}. In massive galaxies, the steeper profiles of ETGs results in them having larger values of the S\'ersic index ($n$ > 2.5), while LTGs typically exhibit indices lower than this value \citep[e.g.][]{Shen2003}. 

While parametric techniques require the assumption of a fitting function (such as a S\'ersic profile), many non-parametric methods have been successfully used to measure morphology from survey data. The pioneering studies of \citet[][C03 hereafter]{Conselice2003}, \citet{Abraham2003} and \citet[][L04 hereafter]{Lotz2004} introduced five such morphological parameters which have been widely used in galaxy morphological classification in the literature: {\color{black} concentration, asymmetry and clumpiness (known collectively as the `CAS' system), \textit{M}$_{20}$ and the Gini coefficient \citep[e.g.][]{Pearson2021}}. Many studies that probe optical and near-infrared images, particularly in the nearby Universe where galaxies are well resolved, show that massive ETGs exhibit higher concentration, lower asymmetry and lower clumpiness than their LTG counterparts \citep[e.g.][]{Conselice2003,Holwerda2014,Cheng2021}. 

Mergers and tidally-disturbed systems typically show the highest values of asymmetry, while galaxies undergoing starbursts typically show the highest values of clumpiness \citep[e.g.][]{Conselice2003}. The Gini -- \textit{M}$_{20}$ space in the optical wavelengths offers a reliable discriminant, at least in the massive-galaxy regime, both between early and late type galaxies and between interacting systems and their non-interacting counterparts \citep[e.g.][]{Lotz2004,Lotz2008}. Similar results have been found in the near-infrared \citep{Holwerda2014}, which is less sensitive to dust obscuration (and therefore a more reliable indicator of the underlying structure of the system), using galaxies from the local Universe in the Spitzer Survey of Stellar Structure in Galaxies (S$^4$G) \citep{Sheth2010}. Forthcoming `Big Data' surveys, from instruments like the Rubin Observatory's LSST \citep{Ivezic2019} and Euclid \citep{Laureijs2010}, will present unique challenges for morphological classification work due to their unprecedented data volumes. {\color{black}Machine learning techniques \citep[e.g.][]{Hocking2018,Martin2020,Walmsley2022,Pearson2022,DominguezSanchez2023}, used in conjunction with visual inspection and morphological parameters, are therefore likely to become important for the next generation of surveys.} 

While a rich literature already exists on galaxy morphology in the massive-galaxy regime \citep[e.g.][]{Michal2020,Fraser-McKelvie2022,Nersesian2023}, much less is known about the dwarf regime (\textit{M}$_\star$ < 10$^{9.5}$ M$_\odot$) outside the very local Universe. For example, are the principal visual morphological classes and their formation histories in dwarf galaxies similar to what is seen in their massive counterparts? How frequent are morphological details like bars and interactions? How do commonly-used morphological parameters perform in the dwarf regime in separating morphological classes and identifying interacting systems? 

The current lack of statistical knowledge about dwarf morphology is partly driven by the fact that, while dwarfs can be studied in detail in the very local Universe, up to distances of $\sim$50 Mpc \citep[e.g.][]{Mateo1998,Tolstoy2009,Besla2016}), typical dwarfs are too faint to be detected, at cosmological distances, in past wide-area surveys like the SDSS \citep[e.g.][]{Jackson2021a,Davis2022}. This is because, while these surveys provide large sky areas, they have relatively shallow detection limits. The dwarfs that are visible in large shallow surveys of the past tend to be those which have relatively high SFRs. The high star formation boosts the luminosity of the dwarfs above the detection thresholds of shallow surveys, making them detectable \citep[e.g.][]{Jackson2021a}. However, this also biases these dwarf samples towards systems that are blue and therefore more likely to have late-type morphology (Kaviraj et al. in prep). {\color{black}Unbiased morphological studies in the dwarf regime have so far been possible, either in our local neighbourhood  \citep[e.g.][]{Tolstoy2009}, in nearby groups and clusters \citep[e.g.][]{Boselli2008,Venhola2017,Eigenthaler2018} or around nearby massive galaxies \citep[e.g.][]{Duc2015,Geha2017,Carlsten2021,Mao2021,Trujillo2021,Holwerda2023}, regions where relatively complete dwarf samples can be assembled. Exploring the morphological mix of dwarfs in the general Universe, in low-density environments, requires surveys which are both deep and wide, like the Hyper Suprime-Cam Subaru Strategic Program \citep[HSC-SSP;][]{Aihara2018a}, in which dwarf galaxy populations are likely to be complete, down to \textit{M}$_{\rm{\star}}$ $\sim$ 10$^{8}$ M$_{\odot}$ out to at least $z\sim0.3$.} 

It is worth noting here that recent work has shown that typical morphological classes, such as ETGs and LTGs, behave differently in the dwarf regime compared to the massive-galaxy regime. For example, \citet{Lazar2023} use ultra-deep HSC-SSP imaging, which has a point source depth of $\sim$28 magnitudes\footnote{The $i$-band 3$\sigma$ limiting surface-brightness (measured in a 10$\times$10 pixel patch) is around 31 mag arcsec$^{-2}$.} (more than 5 magnitudes deeper than standard-depth SDSS imaging) of over 100 blue dwarf ellipticals to show that less than 3 per cent of these systems show signs of tidal interactions, lower than the interaction fraction seen in the general galaxy population at similar stellar masses. This is in contrast to the massive-galaxy regime, in which more than 70 per cent of ellipticals show tidal features in deep images \citep[e.g.][]{vanDokkum2005}. 

The past literature shows that, in the high mass regime, ETGs typically exhibit S\'ersic indices ($n$) around 4, while $n \sim 1$ corresponds to LTGs \citep{Vaucouleurs1959,Vaucouleurs1977,Propris2016}. However, some studies indicates that these trends may not be preserved in the dwarf regime. For example, \cite{Conselice2003}, who study galaxy morphology in the local Universe (out to $\sim$50 Mpc) using optical data, show that concentration cannot be used to separate the ETG and LTG populations in the dwarf regime. Such results suggest that, rather than being strongly shaped by interactions, which would increase the central concentration and result in tidal features, ETGs in the dwarf regime may evolve preferentially via secular processes like gas accretion. More broadly, the evolutionary histories of traditional morphological types (e.g. ETGs and LTGs) in the dwarf regime could potentially be different from those in the massive-galaxy regime.

{\color{black}A detailed census of the morphological mix of dwarf galaxies in the general Universe, outside the local neighbourhood and in low-density environments, is largely missing and clearly desirable for establishing the morphological trends in the dwarf population}. Such a census, which requires an exploration using surveys that are both deep and wide, is the purpose of this study. This paper is organized as follows. In Section \ref{sec:cosmos2020}, we describe the datasets used in this paper and the selection of a sample of nearby dwarf galaxies that underpins this study. In Section \ref{sec:visual}, we explore the morphological mix of our dwarfs via visual inspection of their HSC $gri$ composite images and their unsharp-masked counterparts, identify the principal morphological classes in the dwarf regime, explore the role of morphological details like bars and interactions and study their recent star formation histories via their rest-frame colours. We also compare our findings in the dwarf regime to what is known in massive galaxies. In Section \ref{sec:params}, we explore dwarf morphology using commonly-used morphological parameters, such as the S\'ersic index, concentration, asymmetry and clumpiness (CAS), \textit{M}$_{\rm{20}}$ and the Gini coefficient, and compare the distributions of these parameters in dwarfs to those in their massive counterparts in the nearby Universe. In Section \ref{sec:implications}, we bring our results together and discuss the implications of our findings for the evolution of different dwarf morphological classes. We summarise our findings in Section \ref{sec:summary}. 

%--------------------------------------------------------------

\section{Data}
\label{sec:cosmos2020}

Our study is underpinned by the Classic version of the COSMOS2020 catalogue \citep{Weaver2022}, which provides physical parameters, such as photometric redshifts, stellar masses and SFRs for $\sim$1.7 million sources in the $\sim$2 deg$^2$ COSMOS field \citep{Scoville2007}. These parameters are calculated using deep photometry from 40 broadband filters, across the UV through to the mid-infrared, from the following instruments: GALEX \citep{Zamojski2007}, MegaCam/CFHT \citep{Sawicki2019}, ACS/HST \citep{Leauthaud2007}, Hyper Suprime-Cam \citep{Aihara2019}, Subaru/Suprime-Cam \citep{Taniguchi2007,Taniguchi2015}, VIRCAM/VISTA \citep{McCracken2012} and IRAC/Spitzer \citep{Ashby2013,Steinhardt2014,Ashby2015,Ashby2018}. The optical and infrared aperture photometry is extracted using the \textsc{SExtractor} and \textsc{IRACLEAN} codes respectively and physical parameters are calculated using the \textsc{LePhare} SED-fitting algorithm \citep{Arnouts2002,Ilbert2006}. The 40-filter photometry results in accurate parameters, with photometric redshift accuracies better than $\sim$1 and $\sim$4 per cent for bright ($i<22.5$ mag) and faint ($25<i<27$ mag) galaxies respectively. Our visual inspection uses optical $gri$ colour composite images from the HSC-SSP Ultra-deep layer, which have a 5$\sigma$ point source depth of $\sim$28 magnitudes. This is around 5 magnitudes deeper than standard depth SDSS imaging and almost 10 magnitudes deeper than the detection limit of the SDSS spectroscopic main galaxy sample. The median seeing  of the HSC images is $\sim$0.6 arcseconds (around a factor of 2 better than the SDSS).

%--------------------------------------------------------------

\subsection{A complete sample of nearby dwarf galaxies}
\label{sec:sample}

To construct our dwarf galaxy sample, we first select objects which are classified as galaxies by \textsc{LePhare} (`type' = 0 in the COSMOS2020 catalogue). We then restrict our study to galaxies which have stellar masses in the range 10$^{8}$ M$_{\odot}$ < \textit{M}$_{\rm{\star}}$ < 10$^{9.5}$ M$_{\odot}$ and redshifts in the range $z<0.08$ and which lie within the HSC-SSP footprint and outside the HSC-SSP bright-star masks defined by \citet{Coupon2018}. We restrict ourselves to $z<0.08$ because visual inspection of these dwarfs via their HSC images suggests that morphological classification becomes more difficult beyond this redshift. The sample obtained after applying these constraints is comprised of 283 dwarf galaxies. 

We then further exclude 26 objects that either have an `extendedness' of 0 in the HSC $griz$ filters (i.e. are classified as stars via this parameter, even though they are classified as galaxies by \textsc{LePhare}\footnote{These objects clearly look like stars when visually inspected (see Figure \ref{fig:mimages}).}) or, after visual inspection, either appear to have low signal to noise or are objects where a large foreground or background galaxy is contaminating the photometry of the object in question. The final sample we use for our analysis contains 257 dwarf galaxies. {\color{black}The median redshift and stellar mass errors in our dwarf sample are 0.02 and 0.08 dex, respectively.} Note that $z\sim0.08$ is much lower than the redshift out to which COSMOS2020 is expected to be complete ($z\sim0.3$), for galaxies that have stellar masses of \textit{M}$_{\rm{\star}}$ > 10$^{8}$ M$_{\odot}$ \citep[see e.g. Figure 1 in][]{Jackson2021a}. The sample of dwarfs used here therefore offers an unbiased statistical sample of galaxies which can be used to study the morphological properties of the general dwarf population in the nearby Universe. 

It is instructive to consider the types of large-scale structures that are actually present in the COSMOS2020 footprint at the redshifts probed in this study. Using the X-ray group catalogues compiled by \citet{Finoguenov2007}, \citet{George2011} and \citet{Gozaliasl2019}, which incorporate HST-ACS data from \citet{Leauthaud2007} and photometric redshifts from \citet{Ilbert2009}, we check the \textit{M}$_{200}$ values of groups that reside within the COSMOS2020 footprint at $z<0.1$. The virial masses of the three groups that fit this description lie in the range 10$^{12.9}$ M$_{\odot}$ < \textit{M}$_{\rm{200}}$ < 10$^{13.2}$ M$_{\odot}$. In comparison, a small cluster like Fornax has a virial mass of $\sim$10$^{13.9}$ M$_{\odot}$ \citep{Drinkwater2001}, while larger clusters like Virgo and Coma have virial masses of $\sim$10$^{15}$ M$_{\odot}$ \citep[e.g.][]{Fouque2001,Gavazzi2009}. The COSMOS2020 galaxy population, in our redshift range of interest, therefore resides preferentially in low-density environments.

%--------------------------------------------------------------

\section{Morphological classifications via visual inspection}
\label{sec:visual}

We use visual inspection of HSC $gri$ colour composite images to classify our dwarf galaxies into their principal morphological classes. In order to generate the red-green-blue composite images we make use of the Python function \texttt{make\_lupton\_rgb} \citep[described in][]{Lupton2004} from the python library \texttt{astropy}. Unlike massive galaxies, dwarfs are intrinsically faint and their internal structures (disks, bars, tidal features etc.) may have low contrast in our images. {\color{black}Therefore, for each galaxy, we also create and visually inspect an `unsharp masked' version of each colour composite image.} In unsharp masking \citep[e.g.][]{Malin1977}, a blurred image is created by convolving the original with a kernel. The blurred version is subtracted from the original and the difference image is then multiplied by a factor which represents the strength of the sharpening. The kernel size and sharpening strength are the free parameters in this process. The technique has the effect of sharpening the edges of structures in a galaxy. Given that the sizes of structures may vary in different galaxies, and since our dwarfs span a range (albeit small) in redshift, we explore multiple values for both the blurring kernel and the sharpening strength in each galaxy\footnote{We use the interactive unsharp mask filter in the GIMP software package to interactively vary the kernel and the strength in individual dwarf galaxy images.}.

Unsharp masking has previously been used in astronomy to detect faint, low-contrast features like shells and tidal features inside and around nearby massive galaxies \citep[e.g.][]{Malin1983}. Since they are intrinsically bright, the internal constituents of massive galaxies (spiral arms, bars etc.) are readily visible in contemporary surveys without unsharp masking. However, the same may not be true for dwarfs, which are relatively fainter and unsharp masking could increase the efficacy of our morphological classifications. {\color{black}The visual classifications were performed by one expert classifier (SK). The images were randomised, both the original and unsharp-masked images of each galaxy were classified at the same time and physical parameters (e.g. stellar mass and redshift) were kept hidden during the classification process to avoid introducing any biases}. Our visual inspection records the following information about each galaxy: its morphological class, evidence of an ongoing or recent interaction (tidal features, visible large-scale asymmetry in the main body of the galaxy, shells and dust features) and the presence of a bar. 

%--------------------------------------------------------------

\subsection{Principal morphological classes in dwarf galaxies}

\begin{figure*}
\center
\includegraphics[width=0.88\textwidth]{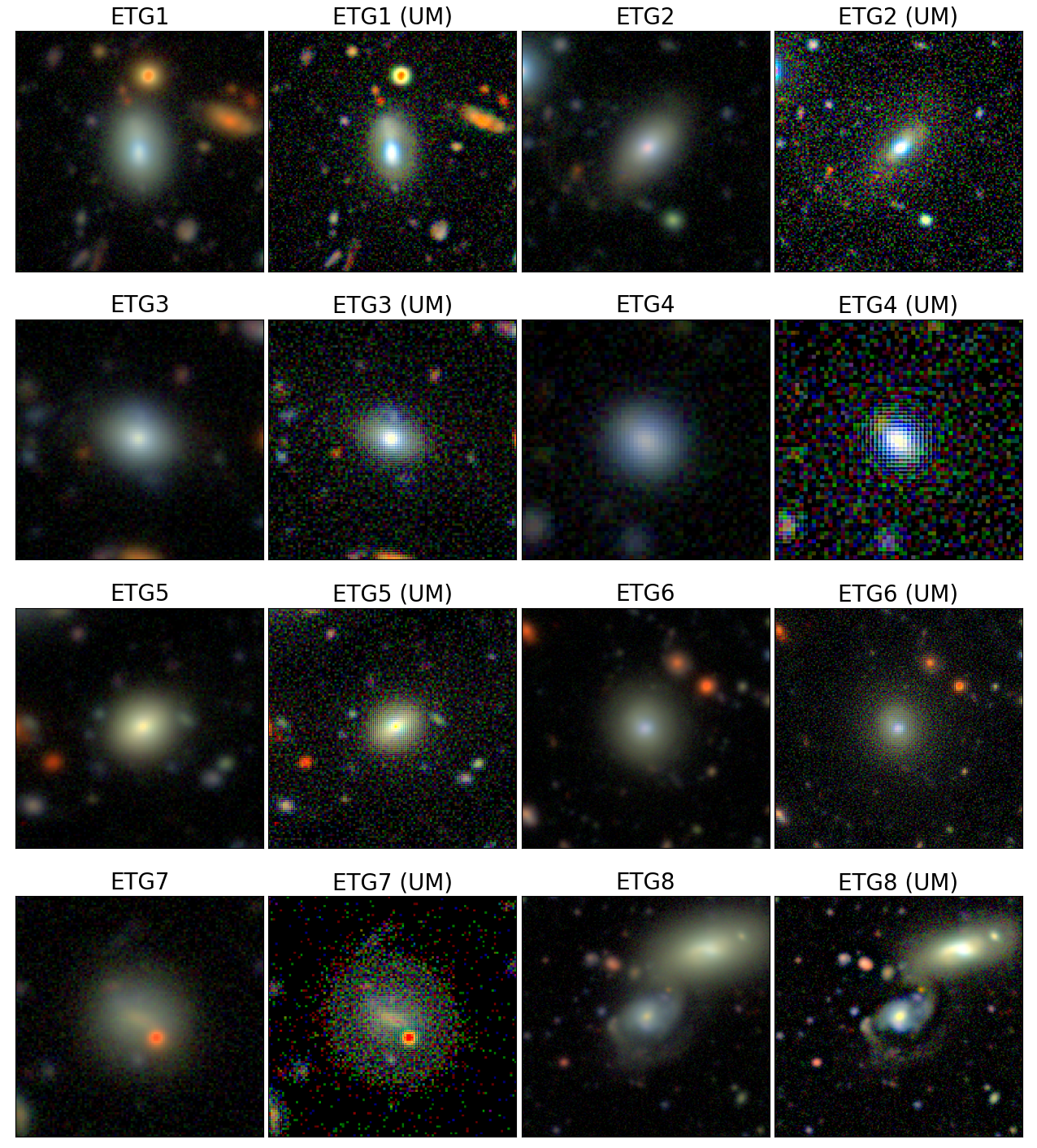}
\caption{{\color{black}Examples of ETGs in our dwarf sample. For each galaxy the left-hand panel shows the HSC colour image, while the right-hand panel shows its unsharp-masked counterpart. ETG1 is an example of a system which shows an internal asymmetry. ETG2 shows a faint tidal feature to the north east of the galaxy. ETG7 is the only dwarf ETG in our sample that exhibits a dust lane, while ETG8 shows an interacting system with two noticeable tidal tails. As the images suggest, blue cores and regions are common in dwarf ETGs whether they are interacting or not.}}
\label{fig:etgimages}
\end{figure*}

\begin{figure*}
\center
\includegraphics[width=0.88\textwidth]{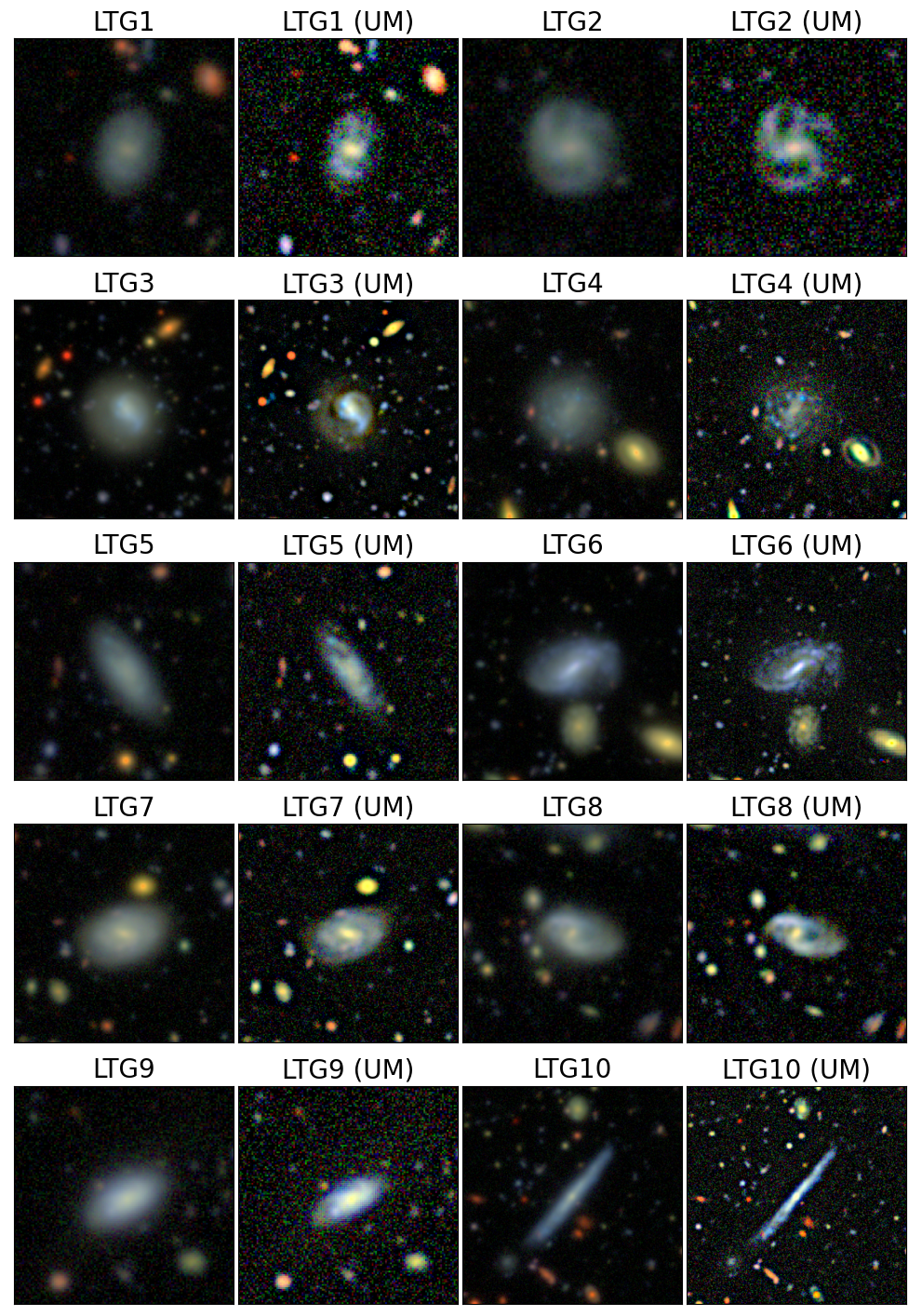}
\caption{Examples of LTGs in our dwarf sample. For each galaxy the left-hand panel shows the HSC colour image, while the right-hand panel shows its unsharp-masked counterpart. LTG3 shows an example of a system which shows an internal asymmetry, while LTG6 shows an example of a system with a tidal feature.}
\label{fig:ltgimages}
\end{figure*}

\begin{figure*}
\center
\includegraphics[width=0.88\textwidth]{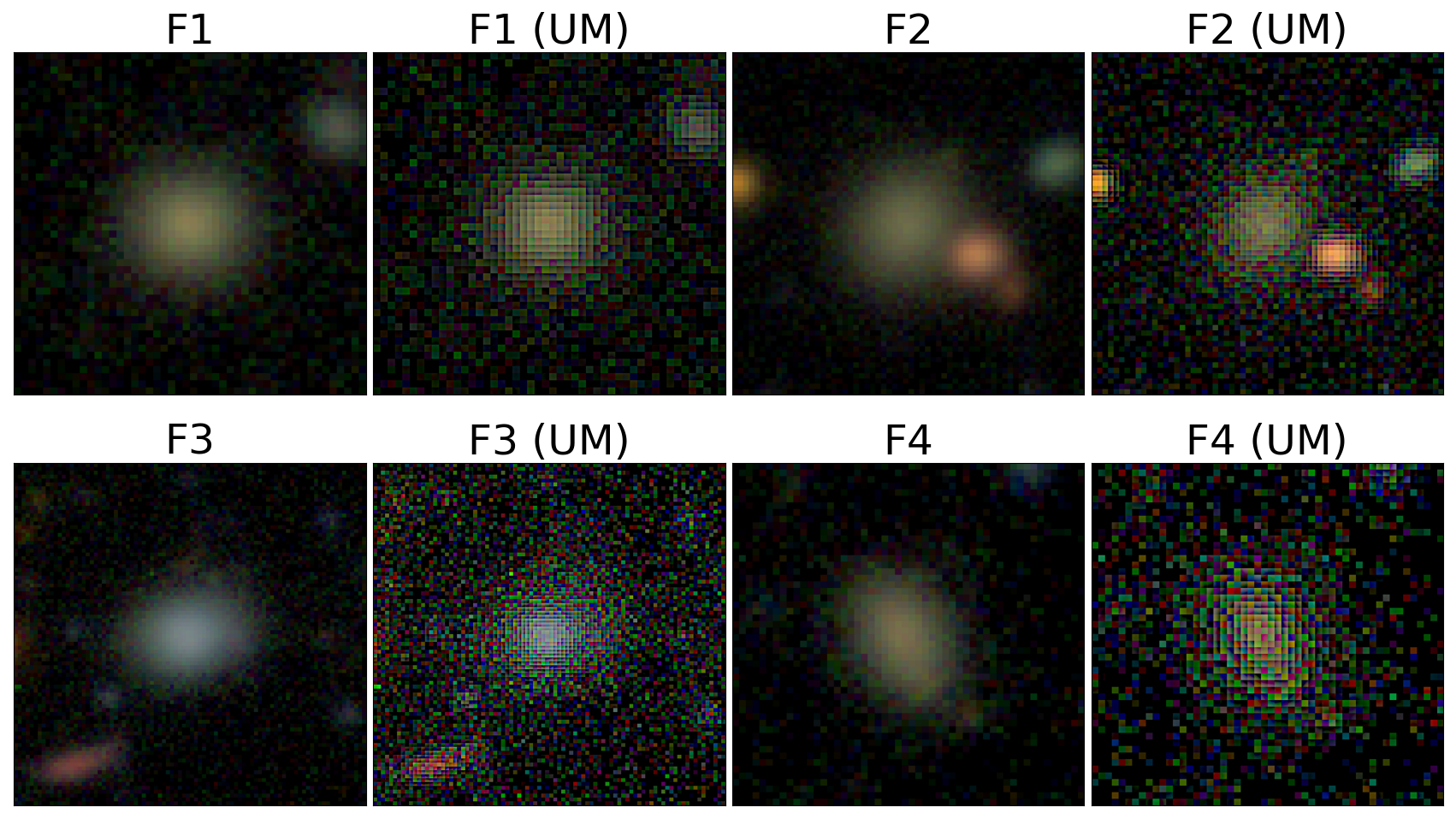}
\caption{Examples of dwarf galaxies classified as featureless. For each galaxy the left-hand panel shows the HSC colour image, while the right-hand panel shows its unsharp-masked counterpart.}
\label{fig:fimages}
\end{figure*}

\begin{figure}
\center
\includegraphics[width=\columnwidth]{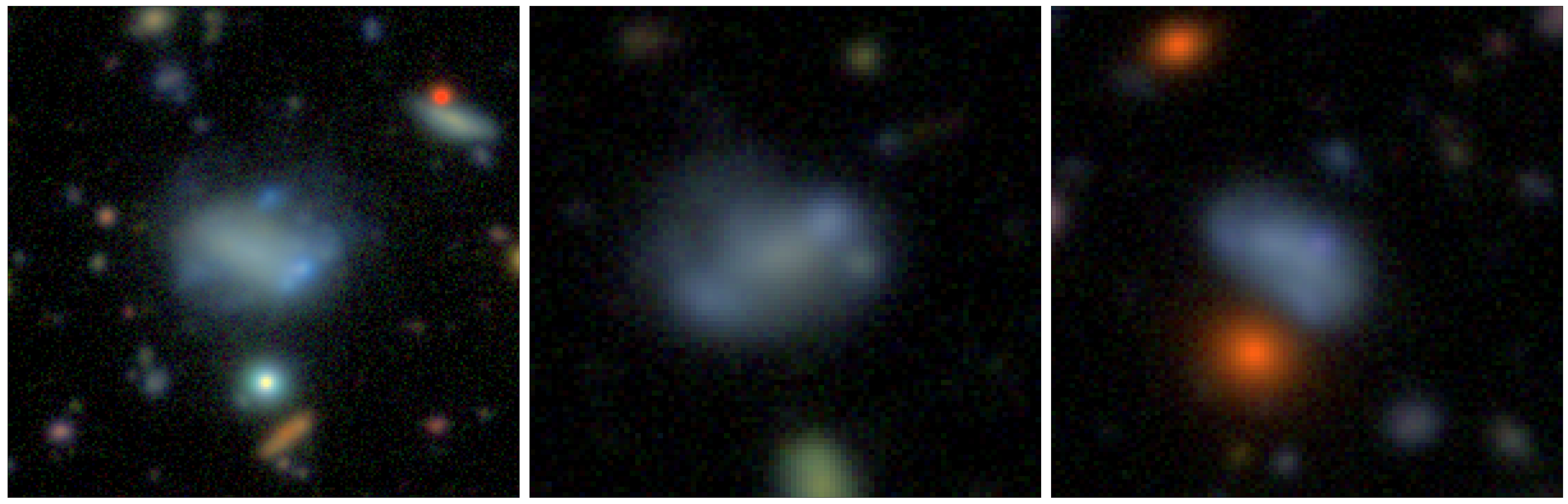}
\caption{Examples of dwarf galaxies which have largely irregular morphology and are difficult to put into the ETG, LTG and featureless categories above. The  fraction of such objects in our sample is negligible (less than 2 per cent).}
\label{fig:uimages}
\end{figure}

{\color{black}We begin by presenting, in Figures \ref{fig:etgimages}, \ref{fig:ltgimages} and \ref{fig:fimages}, the three principal morphological classes yielded by our visual inspection: ETGs, i.e. ellipticals and S0s (Figure \ref{fig:etgimages}), LTGs (Figure \ref{fig:ltgimages}) and featureless dwarfs (Figure \ref{fig:fimages}). ETGs and LTGs in the dwarf regime are similar to the classical morphological types known from studies of massive galaxies. {\color{black}While we do not attempt to divide our LTGs into more granular subclasses, our dwarf LTGs are typically akin to the `Sc' or `Sd' morphologies seen in the massive-galaxy population.} The featureless dwarfs, on the other hand, have flat, smooth profiles that lack either the central light concentration that typifies ETGs or any disk structure that is found in LTGs. A very small number of objects (less than 2 per cent) have a mostly irregular morphology that does not fit into these classes.} Examples of these objects are shown in Figure \ref{fig:uimages}. Given their small contribution to the overall number of dwarfs, these galaxies are omitted from the analysis below for clarity. 

\begin{figure*}
\center
\includegraphics[width=0.88\textwidth]{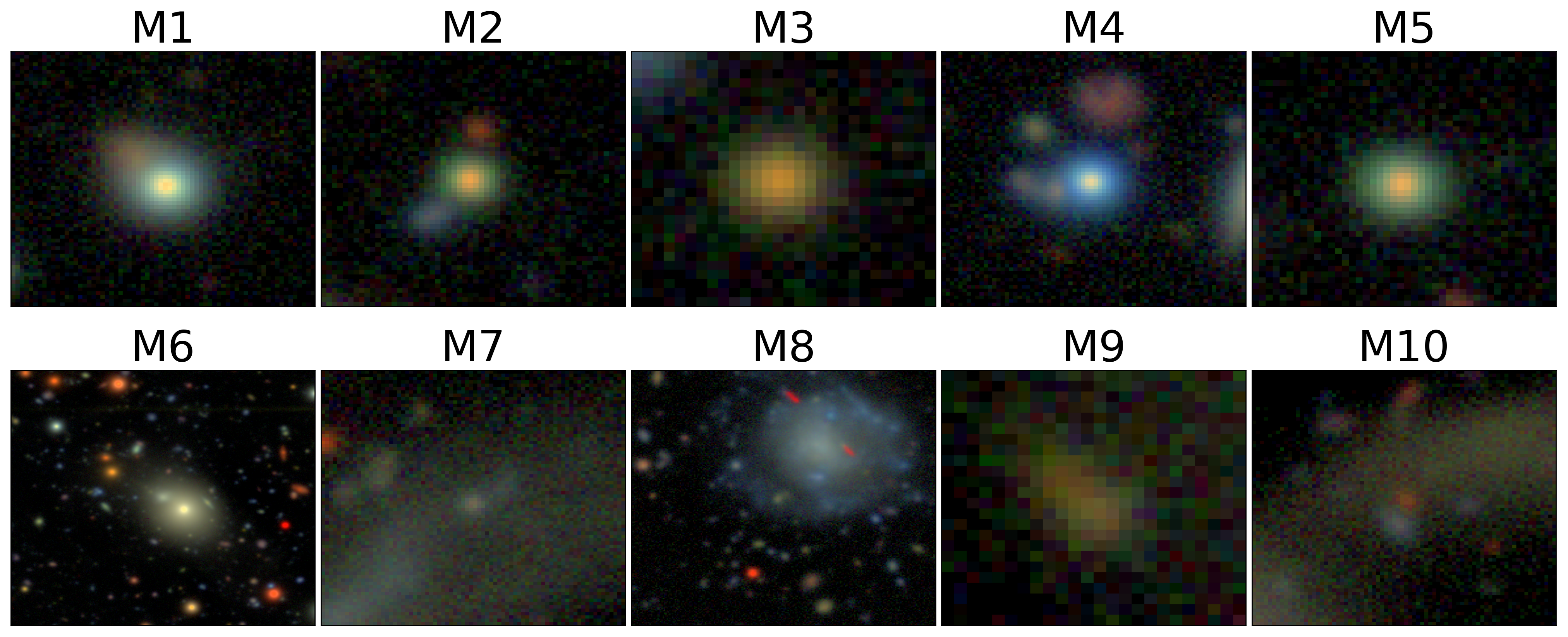}
\caption{Examples of objects that are misclassified and removed from our analysis. These include stars (M1 - 5) which have HSC extendedness values of 0, i.e. are classified as stars by the HSC pipeline even though they are classified as galaxies by LEPHARE. Note that these objects clearly look like stars in the images. M6 and M7 are likely to be contaminated by the light from an overlapping massive galaxy. M8 appears to be an object embedded within the outskirts of a massive galaxy. M9 is a very low S/N object, while M10 is likely to be a high-redshift background galaxy residing behind a broad tidal feature from a nearby galaxy.}
\label{fig:mimages}
\end{figure*}

Finally, in Figure \ref{fig:mimages}, we show examples of the 26 objects that are excluded from our analysis (as described in Section \ref{sec:sample}), either because they are classified as stars by the HSC extendedness parameter (even though they are classified as galaxies by \textsc{LePhare}) or because they are contaminated by nearby massive galaxies or are low signal to noise objects. Table \ref{tab:morphologies} summarises both the number fractions of dwarfs in these different morphological classes (upper sub-table) and the fractions of galaxies in each morphological class which show evidence for ongoing or recent interactions (lower sub-table). The first column (presented in bold) indicates fractions for the entire dwarf sample, while the second and third columns show values for the lower and  upper halves of our stellar mass range respectively. Throughout this paper uncertainties are calculated following \citet{Cameron2011}. 

We note that the featureless class strongly resembles what in the nearby Universe are commonly referred to as dwarf spheroidals (dSph). Indeed, similar galaxies were discovered near the Milky Way as early as the 1930s \citep{shapley38}, and in the Virgo Cluster as early as the 1950s \citep{reaves56, vandenbergh59}.  \citet{reaves56} referred to those in Virgo as IC~3475-type objects, and described them as having "no trace of any small central nucleus or arm structure", with "relatively slight concentration of light to the center, if any", which is extremely similar to our description of the featureless galaxies found in this study. These featureless dwarfs thus may be dSph-type galaxies. 

In this context, it is also instructive to explore the connection between the featureless class and the population of `ultra-diffuse galaxies' \citep[UDGs, e.g.][]{vanDokkum2015,Koda2015}, which are a subset of the dSph population \citep{Conselice2018} and have been prominent in the recent literature. Typical UDGs have effective surface brightnesses ($\mu_{\rm eff}$) fainter than $\sim$25 mag arcsec$^{-2}$ and effective radii ($R_{\rm eff}$) larger than 1.5 kpc \citep[e.g.][]{Conselice2018}. However, while 19 out of 24 galaxies in our featureless class have $R_{\rm eff}$ larger than 1.5 kpc, virtually all of these galaxies have $\mu_{\rm e}$ brighter than 25 mag arcsec$^{-2}$. Only two of our featureless galaxies are consistent with the definition of a UDG (if either $r$ or $i$ band images are considered). Thus, while the featureless objects are likely related to dSPhs, and could be brighter versions of UDG-like systems, we choose to call them featureless for now, until we can conduct more detailed comparisons of their structural properties with dSph galaxies using larger datasets.

%--------------------------------------------------------------

\begin{center}
\begin{table*}
\begin{tabular}{ c | c | c | c }
%\hline
%\hline
\multicolumn{4}{c}{Number fractions in different morphological classes}\\
\toprule
& \textbf{10$^8$ M$_{\odot}$ < \textit{M}$_\star$ < 10$^{9.5}$ M$_{\odot}$} & 10$^8$ M$_{\odot}$ < \textit{M}$_\star$ < 10$^{8.75}$ M$_{\odot}$  & 10$^{8.75}$ M$_{\odot}$ < \textit{M}$_\star$ < 10$^{9.5}$ M$_{\odot}$\\
\midrule
dETG & \textbf{0.43}$^{\textbf{0.03}}$ & 0.45$^{0.03}$ & 0.37$^{0.06}$\\
dLTG & \textbf{0.45}$^{\textbf{0.03}}$ & 0.40$^{0.03}$ & 0.63$^{0.06}$\\
dF & \textbf{0.10}$^{\textbf{0.02}}$ & 0.12$^{0.02}$ & -\\
{\color{black} dIrr} & \textbf{0.02}$^{\textbf{0.01}}$ & 0.03$^{0.01}$ & -\\\\

\multicolumn{4}{c}{Interaction fractions in different morphological classes}\\
\toprule
& \textbf{10$^8$ M$_{\odot}$ < \textit{M}$_\star$ < 10$^{9.5}$ M$_{\odot}$} & 10$^8$ M$_{\odot}$ < \textit{M}$_\star$ < 10$^{8.75}$ M$_{\odot}$  & 10$^{8.75}$ M$_{\odot}$ < \textit{M}$_\star$ < 10$^{9.5}$ M$_{\odot}$\\
\midrule
dETG & \textbf{0.14}$^{\textbf{0.03}}$ & 0.15$^{0.04}$ & 0.10$^{0.06}$\\
dLTG & \textbf{0.28}$^{\textbf{0.04}}$ & 0.24$^{0.04}$ & 0.35$^{0.07}$\\
dF & \textbf{0.20}$^{\textbf{0.07}}$ & 0.20$^{0.07}$ & -\\
{\color{black} dIrr} & \textbf{0.40}$^{\textbf{0.16}}$ & 0.40$^{0.16}$ & -
\end{tabular}
\caption{\textbf{Upper sub-table:} Number fractions of dwarf galaxies in each morphological class (dETG = dwarf early-type galaxies, dLTG = dwarf late-type galaxies, {\color{black} dF = dwarf featureless galaxies, dIrr = dwarf irregular galaxies}), in three mass ranges. The left-hand column shows the number fractions for the full mass range in our sample (10$^8$ M$_{\odot}$ < \textit{M}$_\star$ < 10$^{9.5}$ M$_{\odot}$). In the middle and right-hand columns we split the mass range into its lower and upper halves respectively. Errors, which are shown as superscripts, are calculated following \citet{Cameron2011}. \textbf{Lower sub-table:} The fraction of galaxies in a given morphological class which shows evidence of interactions (see text in Section \ref{sec:visual} for details). {\color{black} The dashes indicate instances where there are no galaxies corresponding to a specific morphological type.}}
\label{tab:morphologies}
\end{table*}
\end{center}

The number fractions of ETGs and LTGs in dwarfs in the general Universe are reasonably similar, although there is an excess of late-types at higher stellar masses. This is different from the situation in high-density environments, reported in other work, where the fraction of dwarf ETGs tends to be higher. For example, the dwarf ETG fraction measured in the MATLAS survey, which samples dwarf satellites around nearby massive galaxies, is $\sim$ 70 per cent \citep{Poulain2021}. Similar results have been obtained by \citet{Carlsten2021}, who performed a comparative study of ETGs and LTGs in the ELVES \citep[][]{Carlsten2022}, NGVS \citep[][]{Ferrarese2012} and NGFS \citep[e.g.][]{Munoz2015} surveys, which probe dwarfs in the local volume, Virgo and Fornax respectively. These trends suggest that the likelihood of dwarfs having early-type morphology increases in regions of higher density, akin to the morphology-density relation seen in the massive-galaxy regime \citep[e.g.][]{Oemler1974,Dressler1980,Sazonova2020}. 

To compare the visual morphological mix of dwarfs to what is known in the massive-galaxy regime, we compare the results of our visual classification to those from  \citet{Kaviraj2010}, who performed morphological classification, via visual inspection, of $\sim$1000 massive galaxies in the SDSS Stripe 82. The galaxies in this sample have upper redshift and lower mass limits of $z\sim0.05$ and \textit{M}$_{\rm{\star}}$ $\sim$ 10$^{10}$ M$_{\odot}$ respectively. Around 60 per cent of the \citet{Kaviraj2010} sample is composed of LTGs, slightly higher than the late-type fraction in the dwarf regime. 

However, the largest difference, in terms of major morphological classes, between dwarfs and massive galaxies is the existence of the featureless galaxies, which do not have an equivalent in the massive-galaxy regime. It is worth noting that all featureless galaxies reside in the lower half of our stellar mass range (see Table \ref{tab:morphologies}). In other words, the fraction of featureless dwarfs increases towards lower stellar masses, which suggests that their formation becomes easier as the gravitational potential well becomes shallower. This, in turn, indicates that their formation likely depends on processes like baryonic feedback or tidal perturbations, whose effect will become stronger as the potential well becomes shallower \citep[e.g.][]{Jackson2021a}.

%--------------------------------------------------------------

\subsection{Local densities of different dwarf morphological classes}

\label{sec:env}

\begin{figure}
    \centering
    \includegraphics[width=0.45\textwidth]{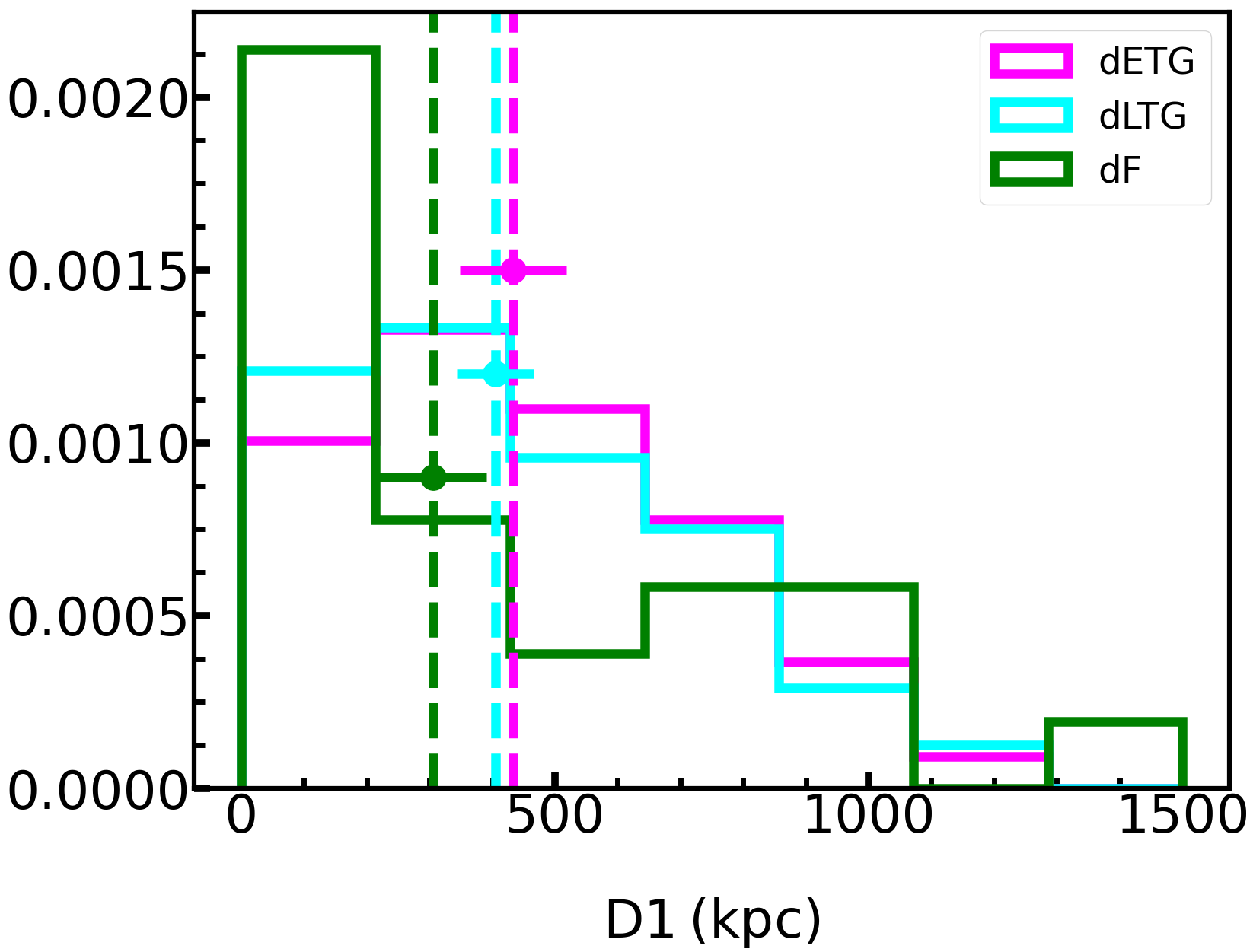} 
    \includegraphics[width=0.45\textwidth]{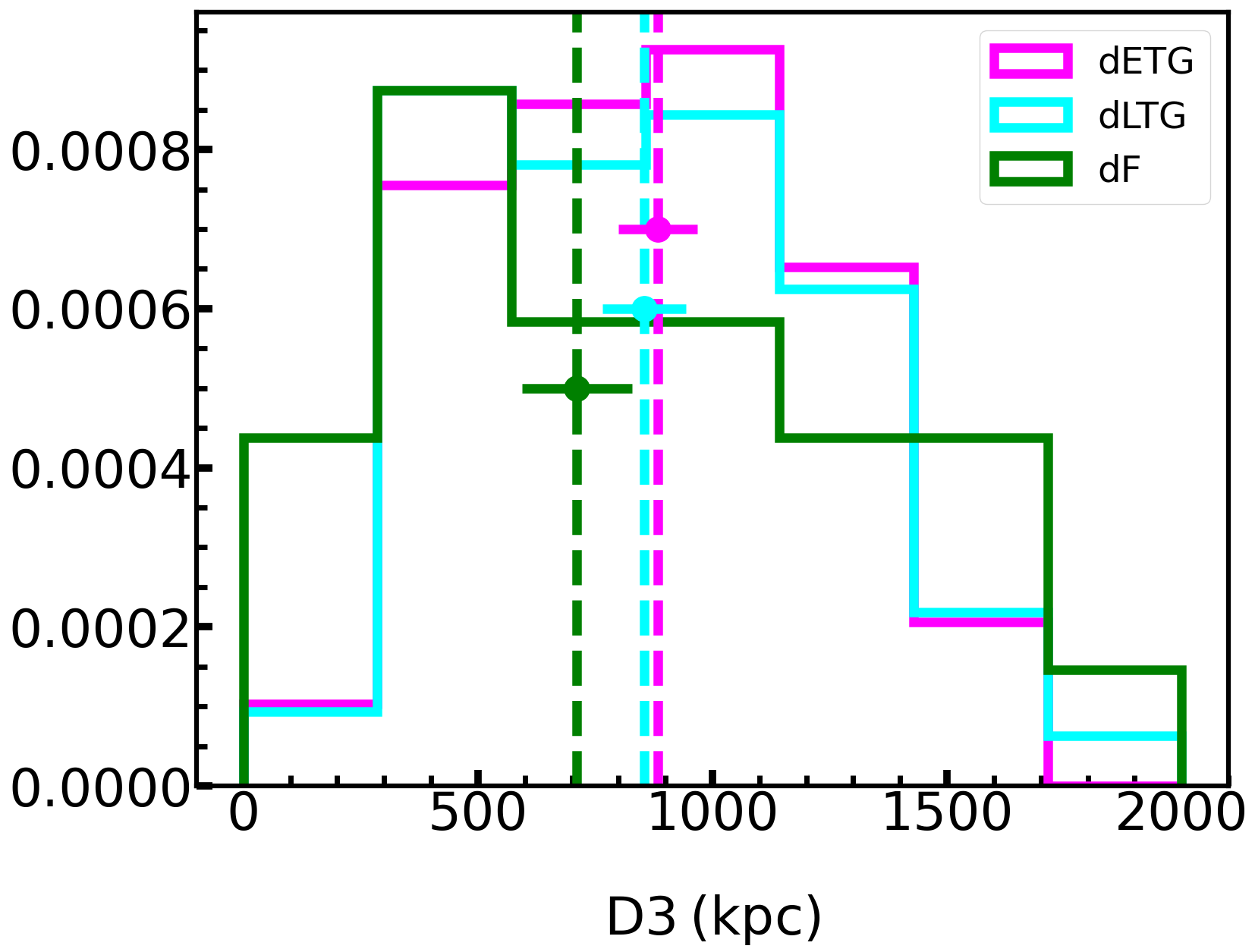}
    \includegraphics[width=0.45\textwidth]{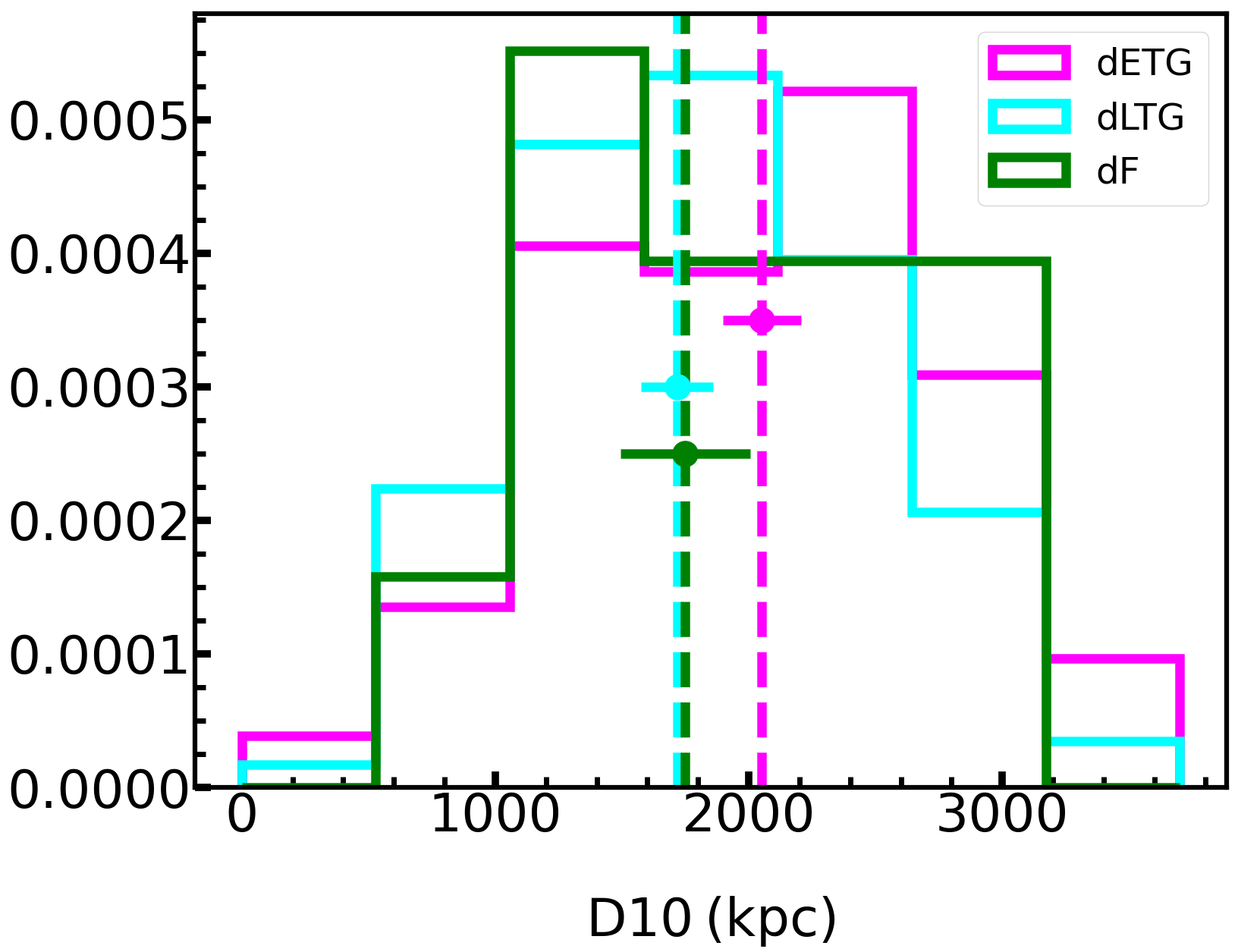}
    \caption{Distributions of projected distances to the first (top panel), third (middle panel) and tenth (bottom panel) nearest massive (\textit{M}$_\star$ > 10$^{10}$ M$_\odot$) neighbour, within a redshift tolerance of 0.02. The histograms are normalized by their areas.}
    \label{fig:dens2}
\end{figure}

We quantify the relative local densities of the various dwarf morphological classes by calculating projected distances of individual dwarfs from the first to the tenth nearest massive (\textit{M}$_\star$ > 10$^{10}$ M$_\odot$) neighbours, using a redshift tolerance of 0.02. While the distance to the first nearest neighbour provides a measure of very local density, distances to progressively farther neighbours provide estimates of the ambient densities averaged over larger distances \citep[e.g.][]{Dressler1980}. Recall that the galaxy population in COSMOS, at the redshifts we are considering here ($z<0.08$), reside in low-density environments (Section \ref{sec:sample}). Our aim here is to understand if there are any strong trends in \textit{relative} density between the different dwarf morphological classes. In Figure \ref{fig:dens2}, we show the distances to the nearest, third nearest and tenth nearest massive neighbour. We find no significant differences between the different dwarf morphological classes in terms of their local density, with the median values overlapping within the statistical errors. The same general trends are present in all other distances (not shown for brevity). It is worth noting, however, that there are more featureless dwarfs at distances less than 250 kpc from the nearest massive galaxies than in other morphological classes (top panel of Figure \ref{fig:dens2}), although this trend has to be confirmed using larger samples of dwarfs in future work.

%--------------------------------------------------------------

\subsection{The role of interactions}

\label{sec:inter}

The interaction fraction in dwarf LTGs ($\sim$28 per cent, Table \ref{tab:morphologies}) is around a factor of 2 higher than that found in massive LTGs, which exhibit interaction fractions between 11 and 16 per cent, depending on the exact morphological subclass \citep{Kaviraj2014}. Recall that our interaction flag includes galaxies that have both tidal features and show large-scale internal asymmetries. The higher interaction fraction in dwarf LTGs, compared to their massive counterparts, is likely to reflect the fact that inducing morphological disturbances in dwarfs is easier (e.g. via mergers, flybys around massive galaxies or internal stellar feedback) due to their shallower potential wells \citep[e.g.][]{Agertz2016,Martin2021}.

{\color{black}Some interesting differences between dwarfs and massive galaxies arise when considering the interaction fraction in dwarf ETGs. Dust lanes in massive ETGs are signposts of merger activity with lower mass companions \citep[e.g.][]{Kaviraj2012,Davis2015}, with around 7 per cent of massive ETGs showing these large-scale dust features \citep{Kaviraj2010}. However, only 1 dwarf ETG (out of 110) shows a dust lane (see Figure \ref{fig:etgimages}). The incidence of dust lanes therefore appears to be markedly lower in dwarf ETGs than in their massive counterparts. In a similar vein, the incidence of interactions in massive ETGs tends to be higher (by around a factor of $\sim$1.5) than in massive LTGs, particularly when compared to the interaction fractions in massive Sc and Sd type systems that are morphologically similar to our dwarf LTGs (see Table 1 in \citet{Kaviraj2014} and \citet{Kaviraj2010}). However, the dwarf regime does not appear to preserve this trend, with dwarf LTGs showing higher interaction fractions, by around a factor of 2, than their ETG counterparts. 

Finally, the interaction fractions in dwarf ETGs are significantly lower than that found in the massive ETGs. For example, while the interaction fraction in dwarf ETGs is $\sim$14 per cent (Table \ref{tab:morphologies}), it is $\sim$70 per cent (i.e. a factor of 5 higher) in massive red ETGs \citep{vanDokkum2005} in images that are around 2 magnitudes shallower than the HSC images used in this study. The interaction fraction in massive galaxies could, therefore, be expected to be even higher in images that have similar depth to the ones used in this study. This suggests that the origin of dwarf ETGs may have less to do with interactions than their massive counterparts, in line with the findings of recent observational work that suggests that secular accretion from the cosmic web is the dominant evolutionary channel for these systems \citep{Lazar2023}. We explore this point in more detail in Section \ref{sec:dwarf_ETG_formation}.}

When large statistical samples are available, the interaction fraction in a population can be combined with the enhancement of the specific SFR (sSFR) in interacting galaxies to estimate the fraction of the star formation activity which is being driven by interactions \citep[Eq. 2 in][]{Kaviraj2014}. 
{\color{black}Large samples likely span the range of properties (e.g. the mass ratio and coalescence timescale of interacting systems) that may influence the interaction-driven sSFR of the system \citep[e.g.][]{Lotz2010a,Lotz2010b}}. Following \citet{Kaviraj2014}, we calculate the enhancement of the sSFR in different dwarf morphological classes by calculating the ratio of the median sSFR in interacting systems to the median sSFR in non-interacting systems. Combining these with the interaction fractions in Table \ref{tab:morphologies} suggests that around 27, 19 and 14 per cent of the star formation activity in our dwarf LTG, featureless and ETG populations are driven by interactions respectively. Combining these values, weighted by the number fractions in the different dwarf morphological classes, suggests that around 20 per cent of the overall star formation activity in dwarfs is triggered by interactions. It is interesting to note that the fraction of star formation activity in massive nearby LTGs that is likely triggered by interactions is higher \citep[around 40 per cent, see][]{Kaviraj2014}. Together with the low frequency of tidal features seen in the dwarf ETGs, this further suggests that, in general, the evolution of dwarf galaxies may have had less to do with interactions than their massive counterparts and that their evolution is likely to be influenced more by secular accretion and internal processes.

\begin{figure}
\center
\includegraphics[width=\columnwidth]{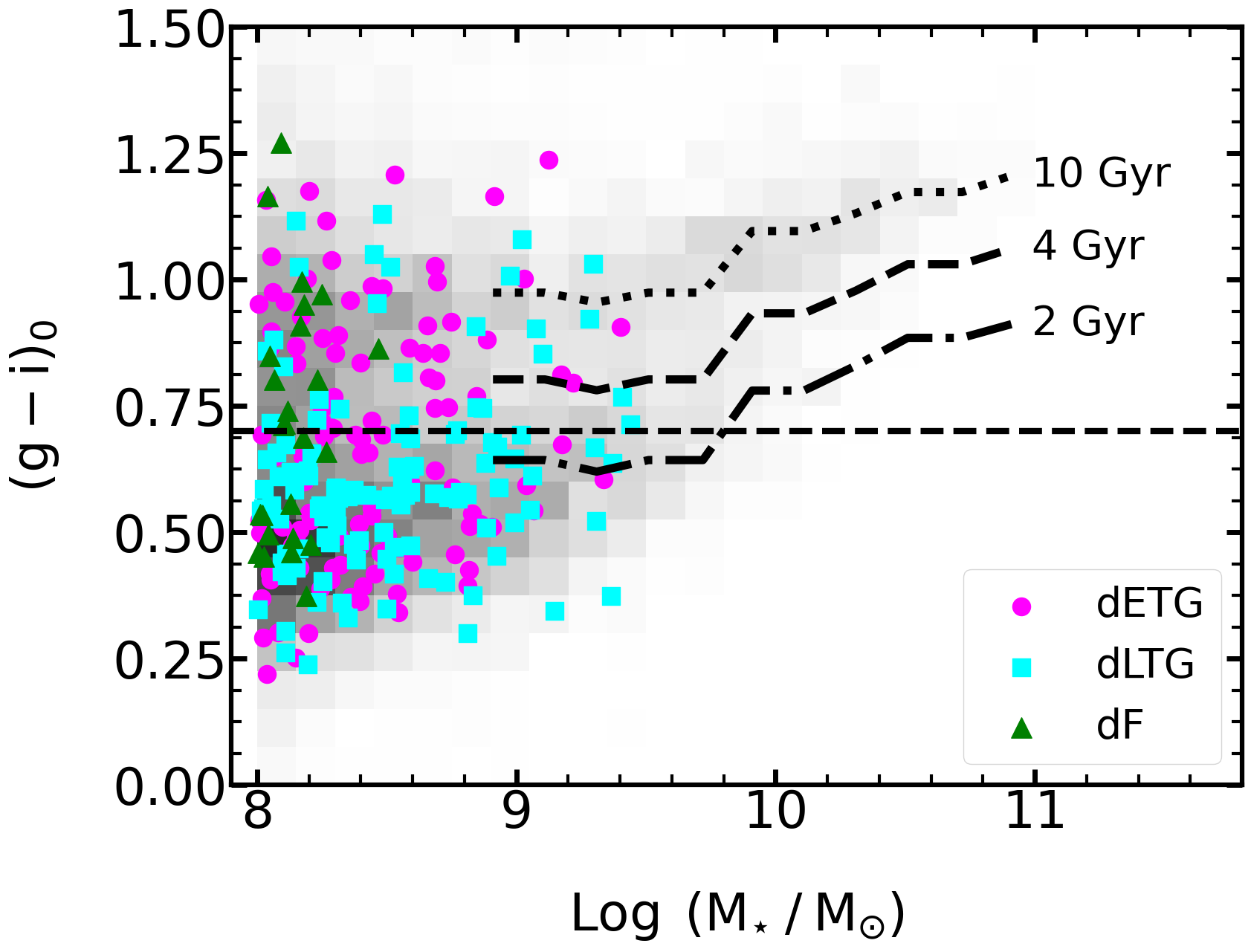}
\includegraphics[width=\columnwidth]{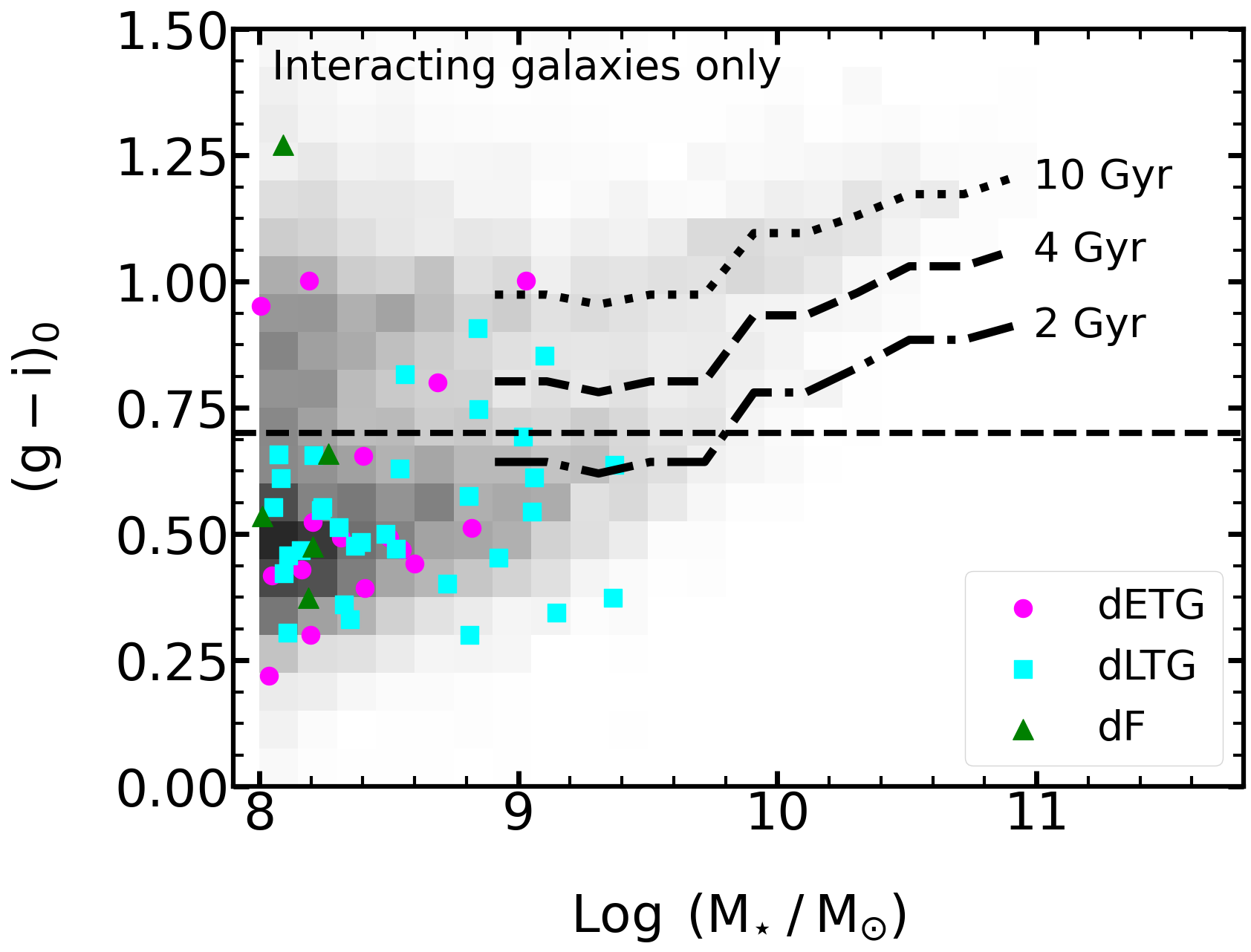}
\caption{\textbf{Top:} Rest-frame $(g-i)$ colours vs stellar mass of COSMOS2020 galaxies at $z<0.2$, shown as a heatmap. The dashed lines show simple stellar populations (SSPs) which form at various look-back times (10, 4, and 2 Gyrs). We use the median metallicity as a function of stellar mass \citet{Gallazzi2005} to construct the rest-frame colour for all SSPs. The three principal dwarf morphological classes are shown overplotted in different colours (see legend). \textbf{Bottom:} Same as the top panel but for interacting galaxies only. The horizontal dashed line at rest-frame $(g-i)=0.7$ separates red and blue galaxies.}
\label{fig:rfgi}
\end{figure}

%--------------------------------------------------------------

\subsection{Rest-frame colours}

\label{sec:colours}

In Figure \ref{fig:rfgi}, we explore the rest-frame colours of the different dwarf morphological classes. The top panel shows the rest-frame $(g-i)$ colour vs stellar mass of the COSMOS2020 population at $z<0.2$ as a heatmap. Galaxies in our dwarf morphological classes are shown overplotted using different colours (see legend). The bottom panel shows the same heatmap as the top panel but this time with only interacting dwarfs overplotted. In the dwarf regime, this heatmap is bimodal around $(g-i)\sim 0.7$ (indicated using the black dashed line), with well-defined red and blue peaks. In both panels we also overlay rest-frame $(g-i)$ colours of simple stellar populations (SSPs) which form at look-back times of 2, 4 and 10 Gyrs (taken from the \citet{Bruzual2003} population synthesis models). Since galaxies show a relationship between stellar mass and metallicity \citep[e.g.][]{Gallazzi2005}, these SSPs use the median stellar metallicity of galaxies at the stellar mass in question. While real galaxies do not have star formation histories that correspond exactly to SSPs, these SSP-based colours serve as a useful guide to understanding the broad features of the star-formation histories of galaxies in this colour-mass diagram. For example, they indicate that a very small fraction of dwarfs (less than 8 per cent in any morphological class) are consistent with purely old stellar populations (i.e. a `monolithic' formation scenario). In a similar vein, dwarfs which reside in the rest-frame $(g-i)$ blue cloud lie blueward of the 4 Gyr SSP, suggesting that they are likely to have had some star formation within the last 4 Gyrs.  

{\color{black} Table \ref{tab:red_fractions} summarises the results in Figure \ref{fig:rfgi}. The upper sub-table shows the red fractions in each morphological class i.e. the fraction of galaxies that have $(g-i)>0.7$, while the lower sub-table shows the same for interacting galaxies only. Together with Figure \ref{fig:rfgi}, we find that the red fractions in interacting dwarfs are lower across all morphological classes, indicating that interactions generally enhance star formation in dwarfs, as is the case in massive galaxies \citep[e.g.][]{Kaviraj2014}. The featureless dwarfs are not all red, which indicates that star formation in at least some of these objects is not completely quenched (or has quenched very recently). Around 60 per cent of dwarf ETGs are blue, and comparison to the 4 Gyr-old SSP suggests that these dwarf ETGs are likely to have had some star formation in the last 4 Gyrs.}

\begin{figure}
\center
\includegraphics[width=\columnwidth]{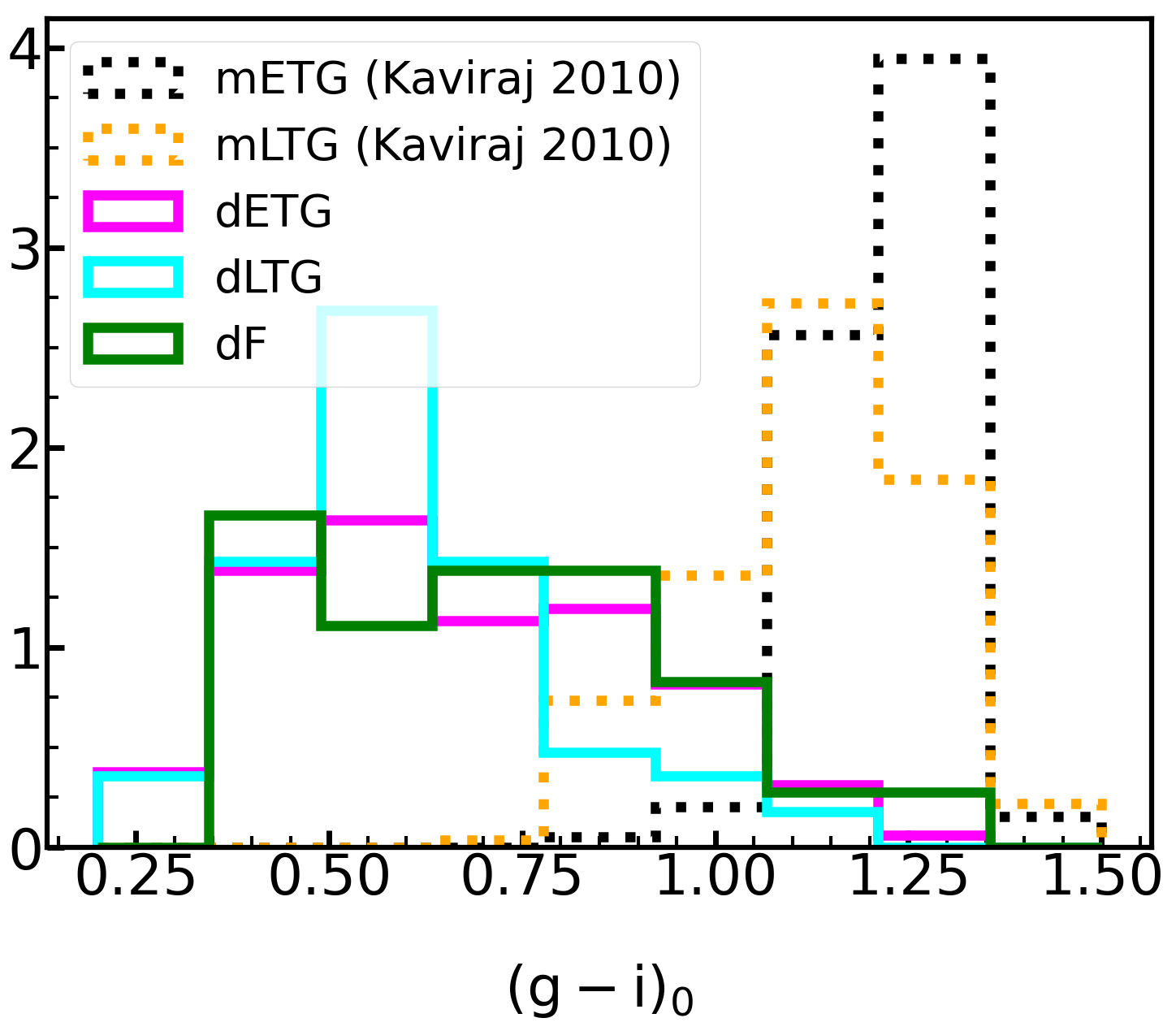}
\caption{Comparison between the rest-frame $(g-i)$ colours of our dwarf morphological classes with their counterparts in the massive-galaxy regime from \citet{Kaviraj2010}.}
\label{fig:rfgi2}
\end{figure}

In Figure \ref{fig:rfgi2}, we compare the colours of our dwarf morphological classes with their counterparts in the massive (\textit{M}$_{\rm{\star}}$ > 10$^{10}$ M$_\odot$) galaxy regime from \citet{Kaviraj2010}. While the majority of dwarf ETGs live in the blue cloud, this is qualitatively different from the massive-galaxy regime, where ETGs tend to be preferentially red. This suggests differences between the formation channels of ETGs in the massive and dwarf regimes, a point we study further in Section \ref{sec:dwarf_ETG_formation}. Nevertheless, the dwarf ETGs do contain the highest fraction of red objects compared to the other morphological classes. 

{\color{black}In it instructive to compare our red fractions  to past work which has probed red/quenched fractions in dwarf galaxies. Some studies, based on the SDSS, have suggested that, at $M_{\star} < 10^9$ \rm M$_{\odot}$, the quenched fraction in nearby ($z<0.055$) dwarfs that reside in low-density environments may be close to zero \citep[e.g.][]{Geha2012}. This is, however, likely to be affected by the selection biases (described in the introduction) whereby red dwarfs preferentially fall out of the selection in shallow surveys outside the local neighbourhood. Indeed, when a similar analysis is performed, using the SDSS, at very low redshift ($z<0.02$), which mitigates some of this bias, the dwarf population shows a pronounced red sequence, which suggests a much higher red/quenched fraction \citep[e.g.][]{Barazza2006}. This agrees with recent work on nearby low-surface-brightness galaxies (which are dominated by dwarfs) using the Dark Energy Survey \citep{Abbott2018}. The red fraction using the $(g-i)$ colour is around 30 per cent (\citet{Tanoglidis2021}, see also \citet{Thuruthipilly2023}), in reasonable agreement with the values we find in this study.
}

\begin{table}
\begin{center}
\begin{tabular}{ c | c }
\multicolumn{2}{c}{Red fractions}\\
\toprule
dETG & 0.41$^{0.04}$\\
dLTG & 0.23$^{0.03}$\\
dF & 0.52$^{0.09}$\\\\

\multicolumn{2}{c}{Red fractions in interacting galaxies}\\
\toprule
dETG & 0.25$^{0.09}$\\
dLTG & 0.13$^{0.05}$\\
dF & 0.20$^{0.14}$
\end{tabular}
\caption{\textbf{Upper sub-table:} Red fractions of dwarf galaxies in different dwarf morphological classes. A galaxy is defined as red if its rest-frame $(g-i)$ colour is greater than 0.7 (see text for details). dETG = dwarf early-types, dLTG = dwarf late-types, dF = dwarf featureless. Errors, which are shown as superscripts, are calculated following \citet{Cameron2011}. \textbf{Lower sub-table:} Red fractions of interacting galaxies in different dwarf morphological classes. }
\label{tab:red_fractions}
\end{center}
\end{table}

\begin{figure*}
\center
\includegraphics[width=0.84\textwidth]{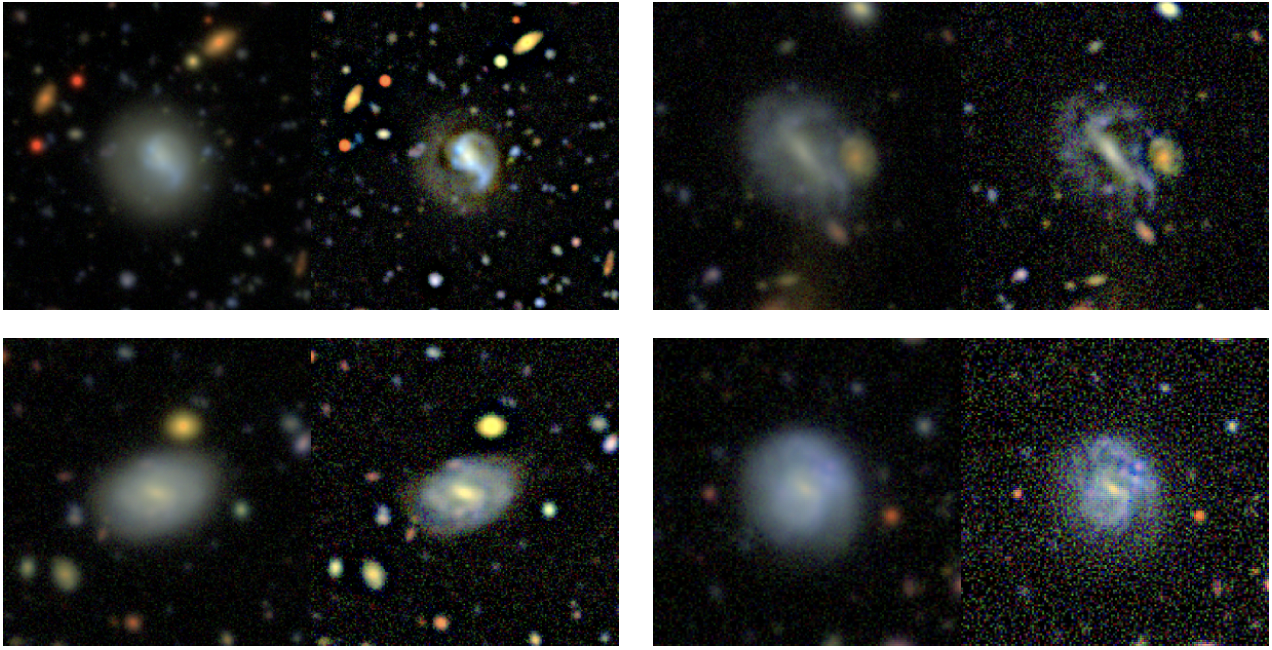}
\caption{Examples of galaxies with bars in our dwarf sample. For each galaxy the left-hand panel shows the HSC colour image, while the right-hand panel shows its unsharp-masked counterpart.}
\label{fig:bars}
\end{figure*}

%--------------------------------------------------------------

\subsection{Bars}

We complete our visual exploration of dwarf morphologies by considering the incidence of bars in our dwarf galaxy sample. Figure \ref{fig:bars} shows examples of four barred galaxies in our dwarfs. The bar fraction in face-on late-type dwarfs is $\sim$11 $\pm$ 3 per cent (9 out of 82 face-on late type galaxies). These are likely to be strong bars \citep[e.g.][]{Nair2010} given that we are using ground-based images of relatively small galaxies. The bar therefore has to extend along a significant fraction of the galaxy to be visible \citep[e.g.][]{Geron2023}. 

The bar fractions at the upper end of the mass range traced by our dwarf galaxies are similar to values reported by recent work \citep[e.g.][]{Geron2021}. In comparison, the strong bar fraction in massive galaxies \citep[e.g.][]{Masters2011,Cheung2013,Geron2021} is around 20 per cent. The frequency of strong bars therefore decreases as we move from the massive-galaxy to the dwarf regime. 

\begin{figure}
\center
\includegraphics[width=\columnwidth]{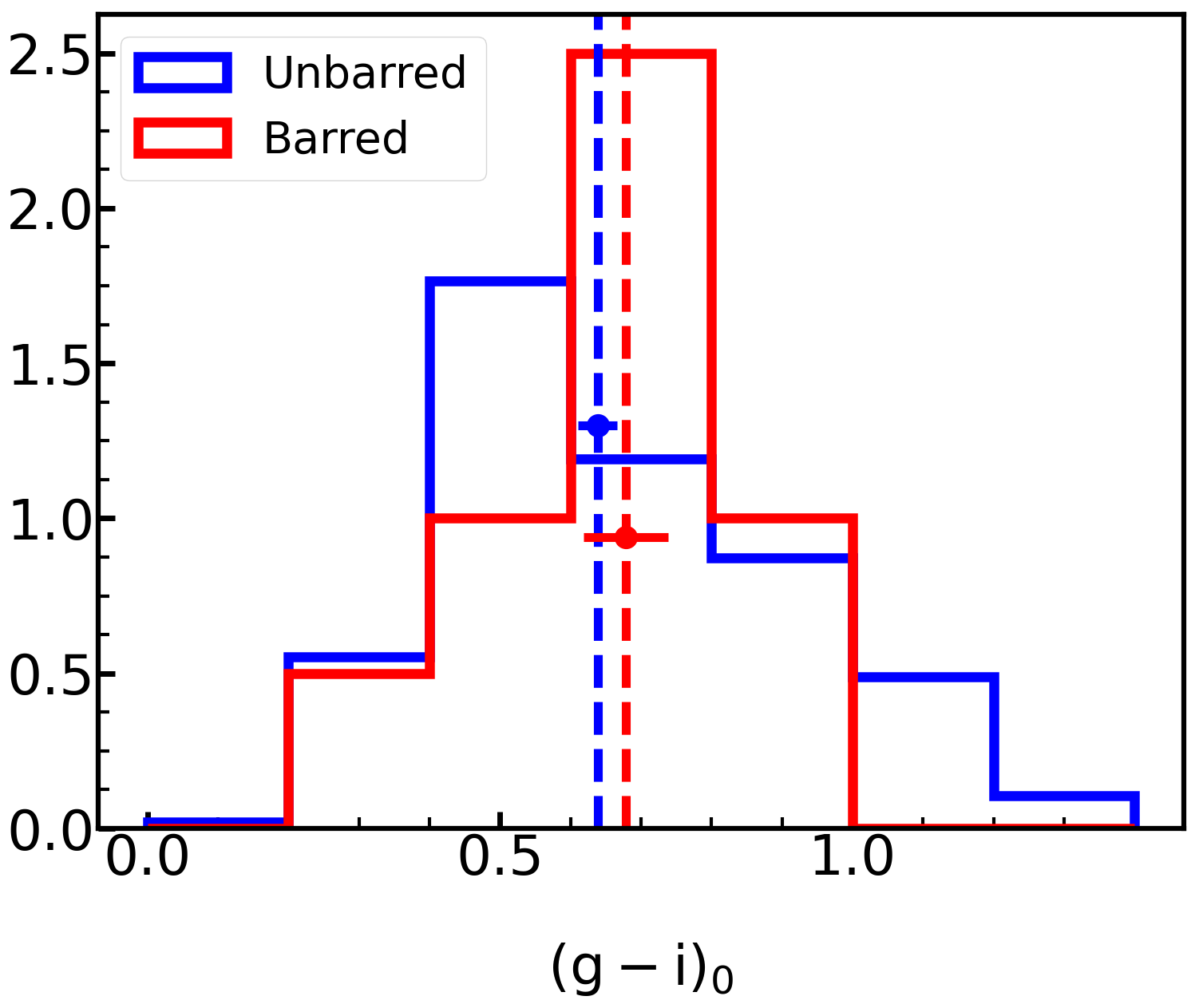}
\caption{Rest-frame $(g-i)$ colours of barred and unbarred dwarfs in our sample. 
}
\label{fig:barcolours}
\end{figure}

Figure \ref{fig:barcolours} shows the rest-frame colours of the barred dwarfs compared to their unbarred counterparts. Barred dwarfs in our sample have a similar colour distribution as unbarred dwarfs with similar median values (which overlap within their uncertainties). A KS test between the two distributions returns a {\color{black} K-S value of 0.22} and p-value of 0.6 indicating that the two distributions are similar. The trends found here appear to be somewhat in contrast to what is found in massive galaxies \citep[e.g.][]{Geron2021} where strongly-barred galaxies tend to be redder.

%--------------------------------------------------------------

\section{Morphological parameters}
\label{sec:params}

We proceed by investigating the differences between dwarf and massive galaxies from the perspective of commonly-used morphological parameters that dominate the recent literature: concentration (C), asymmetry (A), clumpiness (S) (collectively know as the `CAS' system), \textit{M}$_{20}$, the Gini coefficient and the S\'ersic index. We compare our results to previous studies that have used optical data in the dwarf and massive-galaxy regimes, both in the very local Universe (within 50 Mpc) and in the general Universe at low redshift ($z<0.1$).\\

Prior to the calculation of morphological parameters for our dwarf sample we mask any interloper sources, foreground stars and background and foreground galaxies that may interfere with the measurement of the surface brightness profile or morphology of the object in question. In order to obtain accurate unbiased estimates for the morphological parameters studied in this work, we estimate the masked flux using the interpolation algorithm described in \citet{Watkins2022}, which uses the Fourier transform of the azimuthal surface brightness profiles in concentric radial bins to reconstruct the missing flux. {\color{black} The missing flux is reconstructed concentrically until the edge of the image is reached (i.e. where the semi-major axis of the interpolation ellipse reaches the edge of the image) leaving the area outside the interpolation ellipse unchanged. Any masked regions outside the interpolation area are filled with Gaussian noise 1$\sigma$ away from the RMS of the image.} We use the modified interpolated images from the algorithm to calculate effective radii, effective surface brightnesses and central concentrations. The asymmetry, clumpiness, \textit{M}$_{20}$, Gini coefficient and S\'ersic index are estimated by employing the Python \texttt{statmorph} package \citep{Rodriguez-Gomez2019} on the same interpolated images, using Petrosian radii estimated by \texttt{statmorph} as described below. We calculate the morphological parameters and physical properties mentioned above in five HSC bands (\textit{grizy}) for 247 dwarfs from our initial sample of 257. The dwarfs for which parameters cannot be calculated either have relatively low signal to noise, suffer from some light contamination from a nearby bright galaxy or are relatively compact sources.

\subsection{Concentration}
\label{sec:concentration}

We calculate the concentration parameter (\textit{C}$_{82}$), described in Eq. \ref{eq:c82}, as defined by \citet{Kent1985}, \citet{Bershady2000} and C03.

\begin{equation}
    \rm \mathit{C}_{82}=5 \times log_{10} \left( \frac{\mathit{R}_{80}}{\mathit{R}_{20}} \right)
    \label{eq:c82}
\end{equation}

where \textit{R}$_{80}$ and \textit{R}$_{20}$ correspond to the radii enclosing 80\% and 20\% of the total light of the galaxy.

We calculate \textit{R}$\rm_{80}$ and \textit{R}$\rm_{20}$ from the curve of growth, without making any assumptions about the shapes of the light profiles. As opposed to other studies in which set apertures are used, we estimate the total luminosity of galaxies by extrapolating their total luminosities to infinity, where the slope of the curve of growth tends to zero \citep{MunozMateos2015}.

\subsection{Asymmetry}
\label{sec:asymmetry}

The asymmetry parameter as defined by \citet{Abraham1996,Conselice2000,Rodriguez-Gomez2019} is described in Eq. \ref{eq:asy} below. 

\begin{equation}
    \rm \mathit{A}=\frac{\sum_{\mathit{i},\mathit{j}} \mid \mathit{I}_{ij} - \mathit{I}_{ij}^{180} \mid }{\sum_{\mathit{i},\mathit{j}} \mid \mathit{I}_{ij} \mid  } - \mathit{A}_{bgr}
    \label{eq:asy}
\end{equation}

\noindent where \textit{I$\rm_{i,j}$} and \textit{I$\rm_{i,j}^{180}$} are the pixel values of the original and the rotated images, respectively. {\color{black}We calculate asymmetry within a circular aperture of either 1 or 1.5 $\times$ \textit{R}$\rm_{petro}$ depending on the dataset that we are comparing our sample with.} In Sections \ref{sec:local} and \ref{sec:inter2} all pixels within a circular aperture of radius 1.5 $\times$ \textit{R}$\rm_{petro}$ are taken into account for the calculation of asymmetry. However, in Section \ref{sec:outside} the aperture changes to 1 $\times$ \textit{R}$\rm_{petro}$. 
{\color{black} \textit{A}$\rm_{bgr}$ is the asymmetry of the background \citep{Lotz2004}. This quantity is calculated as in Eq. \ref{eq:asy_bgr} where \textit{I$\rm_{bgr;i,j}$} and \textit{I$\rm_{bgr;i,j}^{180}$} are the pixel values of the original and the rotated images respectively, corresponding to the region outside the circular aperture, which we define as the background.}

{\color{black}
\begin{equation}
    \rm \mathit{A_{\rm bgr}}=\frac{\sum_{\mathit{i},\mathit{j}} \mid \mathit{I}_{bgr;ij} - \mathit{I}_{bgr;ij}^{180} \mid }{\sum_{\mathit{i},\mathit{j}} \mid \mathit{I}_{bgr;ij} \mid}
    \label{eq:asy_bgr}
\end{equation}
}

\subsection{Clumpiness}
\label{sec:clumpiness}

The clumpiness (sometimes also referred to as the smoothness index), as defined in L04 and \citet{Rodriguez-Gomez2019}, is described in Eq. \ref{eq:s} below.

\begin{equation}
    \rm \mathit{S}=\frac{\sum_{\mathit{i},\mathit{j}}  \mathit{I}_{ij} - \mathit{I}_{ij}^{S} }{\sum_{\mathit{i},\mathit{j}} \mathit{I}_{ij}} - \mathit{S}_{bgr}
    \label{eq:s}
\end{equation}

\noindent where \textit{I}$\rm_{i,j}$ and \textit{I}$\rm_{i,j}^{S}$ are the pixel values of the original image and its smoothed version, respectively, {\color{black} within circular apertures of 1 or 1.5 $\times$ \textit{R}$\rm_{petro}$, depending on which dataset we are comparing our sample with.} The smoothed image is obtained using a boxcar filter of width $\sigma$, which is set to 0.25 $\times$ \textit{R}$\rm_{petro}$, as in \citet[][L04 hereafter]{Lotz2004}. The calculation is performed only for the pixels corresponding to radii between $\sigma$ and 1.5 $\times$ \textit{R}$\rm_{petro}$ {\color{black} (or 1 $\times$ \textit{R}$\rm_{petro}$ for the comparison with DES data in Section \ref{sec:outside})}, since the central region is avoided due to most galaxies showing significant central concentration. The pixels with a negative numerator value in Eq. \ref{eq:s} are excluded from the summation. S$\rm_{bgr}$ is the background clumpiness, calculated using the background pixels residing outside the segmentation map (Eq. \ref{eq:sbgr}).

\begin{equation}
    \rm \mathit{S\rm_{bgr}}=\frac{\sum_{\mathit{i},\mathit{j}}  \mathit{I}_{bgr,ij} - \mathit{I}_{bgr,ij}^{S} }{\sum_{\mathit{i},\mathit{j}} \mathit{I}_{bgr,ij}}
    \label{eq:sbgr}
\end{equation}

\noindent where \textit{I$\rm_{bgr;i,j}$} and \textit{I$\rm_{bgr;i,j}^{S}$} are the pixel values of the original image and its smoothed version, respectively, corresponding to the region outside the circular aperture, which we define as the background.

\subsection{Gini coefficient}
\label{sec:gini}

We calculate the Gini coefficient \citep{Abraham2003,Lotz2004,Rodriguez-Gomez2019} as described in Eq. \ref{eq:gini}. 

\begin{equation}
    \rm \mathit{G}=\frac{1}{\mid\mathit{\overline{X}}\mid \mathit{n} (\mathit{n}-1)} \sum_{\mathit{i}=1}^{\mathit{n}} (2\mathit{i}-\mathit{n}-1) \mid \mathit{X}_i \mid
    \label{eq:gini}
\end{equation}

\noindent where \textit{n} is the number of pixels in the image, $\overline{X}$ is the mean of the pixel values and \textit{X}$\rm_i$ corresponds to the value of each pixel. Following \citet{Lotz2004} we use a segmentation map for the calculation of the Gini coefficient. The segmentation map is constructed by convolving the image with a Gaussian kernel of $\sigma$ = \textit{R}$\rm_{petro}$/5. We then select pixels from the original image that are above the mean surface brightness of the convolved image at \textit{R}$\rm_{petro}$.  

\begin{figure}
\center
\includegraphics[width=0.9\columnwidth]{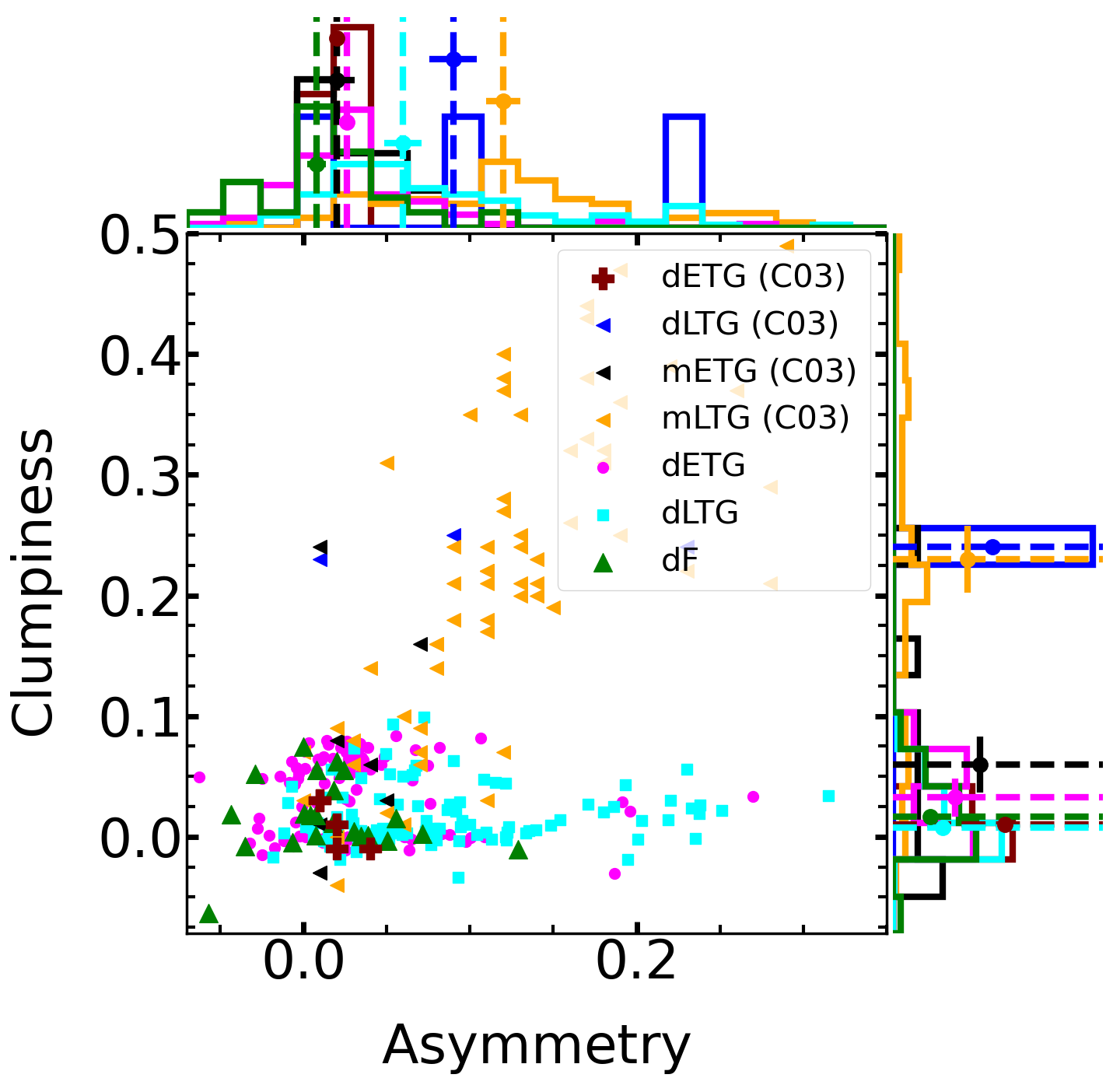}\\ 
\includegraphics[width=0.9\columnwidth]{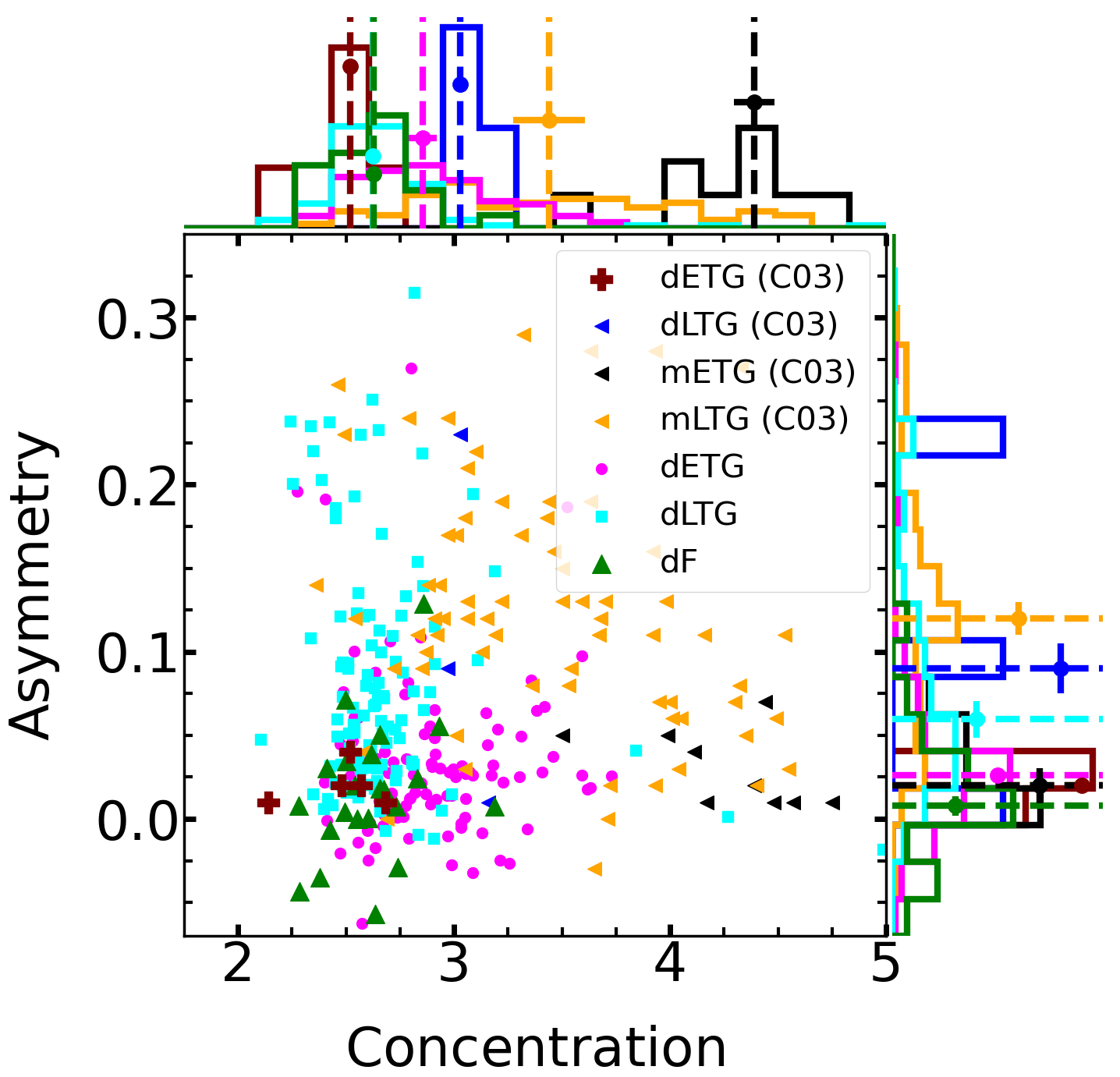}\\
\includegraphics[width=0.9\columnwidth]{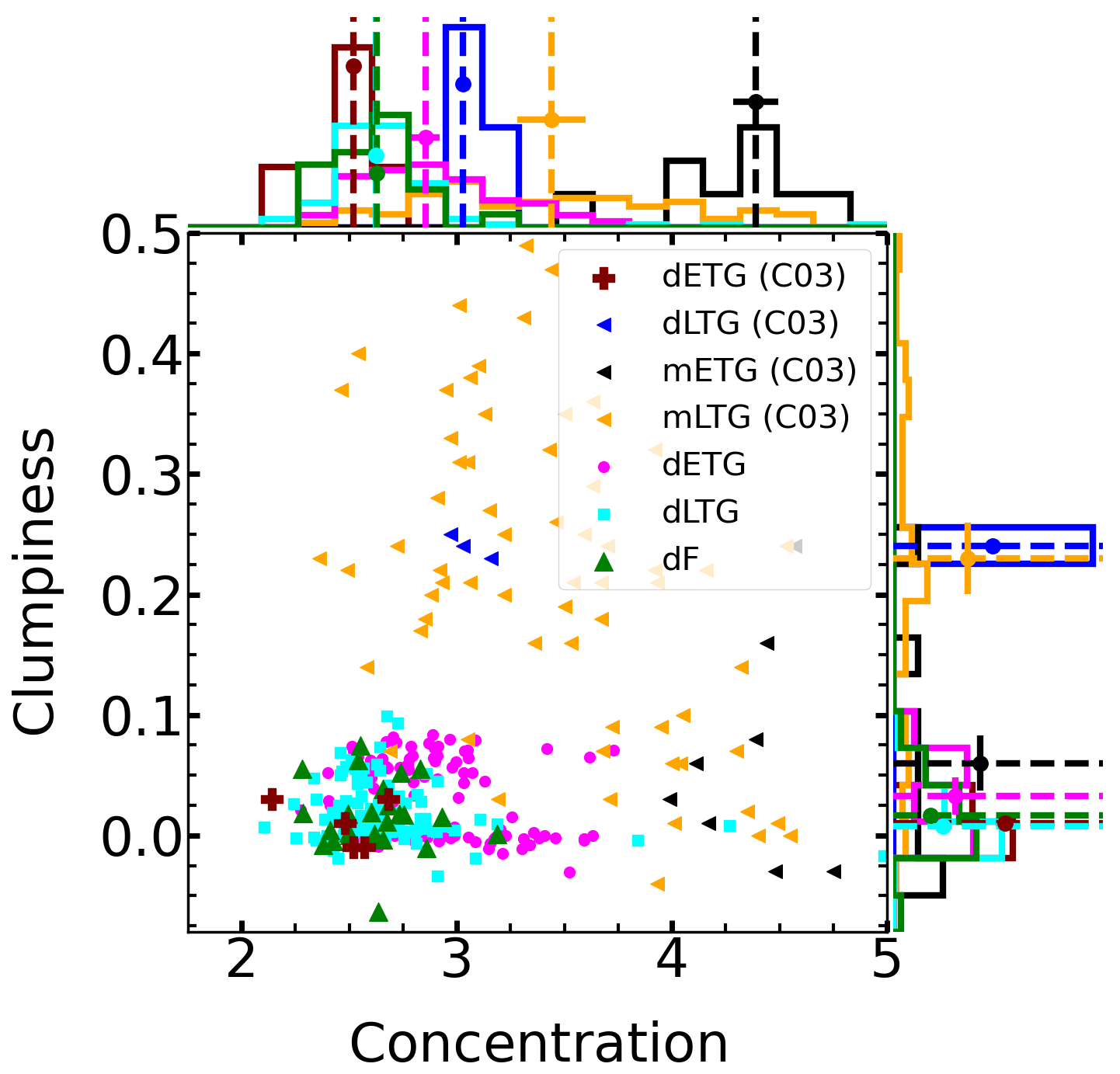}
\caption{Combinations of CAS parameters for various morphological classes in our dwarf and the C03 samples. Distributions of parameters and median values (together with their associated uncertainties) are shown on the sides of all panels. `m' and `d' correspond to the massive (\textit{M}$_\star$ > 10$^{10}$ M$_\odot$) and dwarf (10$^{8}$ M$_\odot$ < \textit{M}$_\star$ < 10$^{9.5}$ M$_\odot$) regimes, respectively. ETG = early-type galaxy, LTG = late-type galaxy, F = featureless galaxy.}
\label{fig:aVSs_local}
\end{figure}

\subsection{\textit{M}$_{20}$}
\label{sec:m20}

{\color{black} The \textit{M}$_{20}$ index (L04) is the second order moment of the brightest 20 per cent of the galaxy's flux, divided by the total second order central moment ($\rm \textit{M}_{tot}$; see Eq. \ref{eq:mu}).} The formula for calculating this statistic is given in Eq. \ref{eq:m20}.

\begin{equation}
    \rm \mathit{M}_{tot}=\sum_\mathit{i}^\mathit{n} \mathit{M}_i=\sum_\mathit{i}^\mathit{n} \mathit{f}_i ((\mathit{x}_i-\mathit{x}_c)^2+(\mathit{y}_i-\mathit{y}_c)^2)
    \label{eq:mu}
\end{equation}

\noindent where \textit{f}$\rm_i$ is the flux of each pixel, \textit{x}$\rm_i$ and \textit{y}$\rm_i$ represent the positions of each pixel, \textit{x}$\rm_c$ and \textit{y}$\rm_c$ correspond to the locations of the galaxy's centre, which are estimated by minimizing \textit{M}$\rm_{tot}$.

\begin{equation}
    \rm \mathit{M}_{20}=log_{10}\left(\frac{\sum_{i,BRG} \mathit{M}_i}{\mathit{M}_{tot}}\right)
    \label{eq:m20}
\end{equation}

\noindent where $\rm\sum_{i,BRG}$ \textit{M}$\rm_i$ represents the summation over the brightest pixels of the image corresponding to 20 per cent of the galaxy's total flux. The pixels which are taken into account for the calculation of this statistic are selected using the Gini segmentation map described in Section \ref{sec:gini}. {\color{black} We note that the segmentation maps (i.e. the collections of pixels taken into account) used to calculate concentration, asymmetry and clumpiness are different, because they are either defined based on set apertures using \textit{R}$\rm_{petro}$ or by extrapolating the total galaxy luminosity to infinity (in the case of concentration).}

\subsection{S\'ersic index}
\label{sec:S\'ersic_index}

The details of the S\'ersic fitting procedure in this study can be found in \citet{Rodriguez-Gomez2019} as we use the \texttt{statmorph} package to perform the fitting. We fit 2D single component S\'ersic profiles using the Python \texttt{astropy} modelling package within \texttt{statmorph}. 

In the following sections we compare the morphological parameters derived for our dwarf galaxies first to dwarfs and massive galaxies in the very local Universe (within 50 Mpc) and then to massive galaxies in the general Universe at low redshift ($z<0.1$). We then consider whether combinations of parameters exist which can separate interacting dwarfs from their non-interacting counterparts. We perform these comparisons via parameters calculated using HSC \textit{r} band images, which is closest to the filters used in previous studies. 

%--------------------------------------------------------------

\subsection{Comparison to dwarf and massive galaxies in the very local Universe (within 50 Mpc)}

\label{sec:local}

\begin{figure}
\center
\includegraphics[width=\columnwidth]{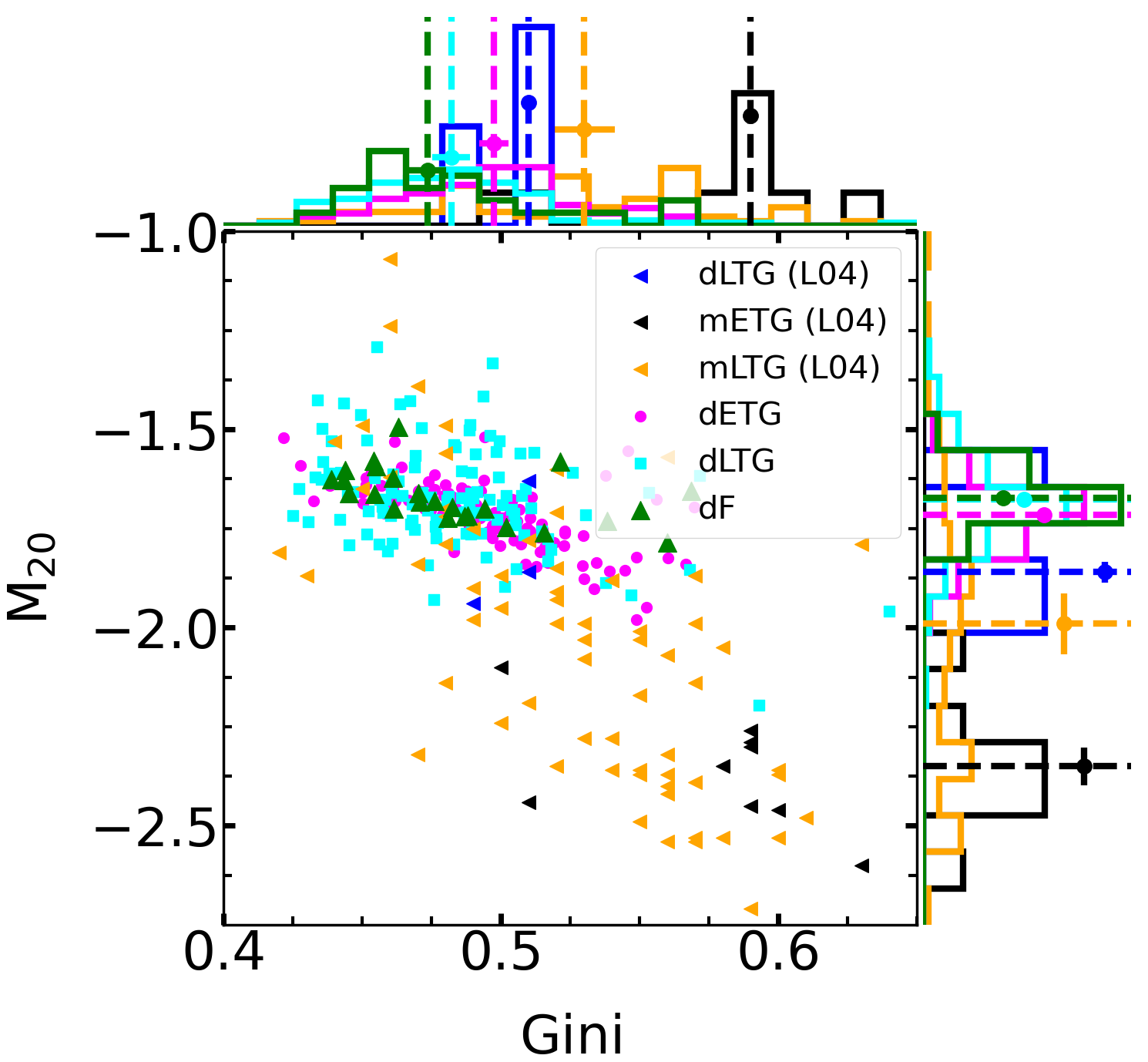}
\caption{M$_{20}$ vs Gini coefficient for various morphological classes in our dwarf and the L04 samples. Distributions of parameters and median values (together with their associated uncertainties) are shown on the sides of all panels. `m' and `d' correspond to the massive (\textit{M}$_\star$ > 10$^{10}$ M$_\odot$) and dwarf (10$^{8}$ M$_\odot$ < \textit{M}$_\star$ < 10$^{9.5}$ M$_\odot$) regimes, respectively. ETG = early-type galaxy, LTG = late-type galaxy, F = featureless galaxy.}
\label{fig:m20VSg_local}
\end{figure}

In Figures \ref{fig:aVSs_local} and \ref{fig:m20VSg_local} we compare our results to C03 and L04 who have used the CAS system, \textit{M}$_{20}$ and the Gini coefficient to study galaxy morphology in the optical wavelengths in the very local Universe (within 50 Mpc), in galaxies that span all Hubble types from \citet{Frei1996}. It is worth noting that the numbers of galaxies in these studies is relatively small, which may affect the reliability of the comparisons. However, the proximity of the galaxies studied by C03 and L04 means that the dwarf samples are relatively unbiased in terms of morphological type, stellar mass and colour distribution, although there is still a bias because the dwarfs from these studies reside in a relatively dense environment. Nevertheless, such a comparison can provide a useful sanity check of the parameter values we have found in our dwarf population. {\color{black} In order to perform this comparison consistently with these previous studies, we consider asymmetry and clumpiness for our dwarfs calculted using pixels that reside within 1.5 $\times$ \textit{R}$\rm_{petro}$, as in C03 and L04.} We also restrict dwarfs from C03 and L04 to the stellar mass range spanned by our HSC dwarfs. 

It is instructive to note the physical scales that are being traced in the images used to calculate morphological parameters in the different datasets. {\color{black}Given the upper redshift limit of our dwarf sample ($z\sim0.08$) and a median HSC seeing of 0.6 arcseconds, the physical scales resolved in these galaxies is $\sim$0.9 kpc or better. Given a median seeing of around 2 arcseconds on the Palomar P60 telescope, the Frei et al. images are able to resolve physical scales of $\sim$0.4 kpc or better in these galaxies (within a factor of 2 of our HSC dwarf sample).}

The top panel of Figure \ref{fig:aVSs_local} presents the clumpiness vs asymmetry plane for both our dwarfs and the C03 galaxies. The distributions of parameters and median values (together with their associated uncertainties) are shown on the sides of all panels. It is worth noting first that, while the different morphological classes within our dwarf sample show differences in asymmetry, they show significant overlap in clumpiness. The dwarf featureless galaxies show the lowest median value of asymmetry. After that, the galaxies with the lowest asymmetries are dwarf and massive ETGs, while those with the highest asymmetries are late-type systems \citep[consistent with what is seen in other work e.g.][]{Bershady2000,TaylorMager2007}. 

Our dwarf ETGs reside in similar regions of this plane as their C03 counterparts. Our dwarf LTGs and the massive LTGs from C03 show reasonable overlap in asymmetry. However our dwarf LTGs show a significant offset in the median values of clumpiness, with both dwarf and massive LTGs from C03 (the latter having clumpiness values which are a factor of $\sim$5 higher than our dwarf LTGs). The difference between the dwarf LTGs in our sample and those in C03 could be driven by the fact that the local Universe galaxies are better resolved, although the very small numbers make it difficult to come to a definitive conclusion. A potential reason for the significant difference in clumpiness between dwarf and massive LTGs is that the SFRs are higher in the massive-galaxy regime, where the potential wells are steeper and gas accretion takes place more efficiently. This is likely to cause the galactic structure to contain more HII regions which, in turn, would tend to increase the `patchiness' (or clumpiness) of the system. {\color{black}Interestingly, the same differences are seen when we compare our dwarfs to massive galaxies from the DES survey, which are resolved slightly worse than our HSC dwarfs (see Section \ref{sec:outside} below), suggesting that the trend may indeed be driven by differences in physical properties, such as SFR, than simply the resolution of the images.}

The middle and bottom panels of Figure \ref{fig:aVSs_local} present the asymmetry vs concentration and the clumpiness vs concentration planes. The concentrations of our dwarf ETGs are similar to those measured in their counterparts from C03. {\color{black} However, massive ETGs show, on average, higher concentration values than the dwarf ETGs (both from our sample and C03) by a factor of $\sim$1.7.} As noted already in C03, the structural differences between dwarf and massive ETGs suggest that they form via different evolutionary channels. The observed differences in terms of morphological parameters appear aligned with the results of Section \ref{sec:inter}, which demonstrates that dwarf ETGs appear to exhibit a lower incidence of interactions than their high mass counterparts. Thus, while massive ETGs are known to experience minor and major merger events, which are likely to increase the concentration of these systems \citep[e.g.][]{Martin2018_sph}, the lower concentrations and lack of interaction signatures in dwarf ETGs suggest an evolutionary history that is dominated by secular accretion. We explore this point further in Section \ref{sec:dwarf_ETG_formation}.

In Figure \ref{fig:m20VSg_local}, we complete the comparison of our dwarfs to galaxies in the very local Universe by considering the \textit{M}$_{20}$ -- Gini plane. L04 demonstrate that, in the massive-galaxy regime, ETGs reside at relatively high and low values of Gini and \textit{M}$_{20}$ respectively, while the opposite is true for the LTG population. In other words, these morphological classes are reasonably well separated in the \textit{M}$_{20}$ -- Gini plane. However, this figure indicates that dwarf ETGs and LTGs show significant overlap in the \textit{M}$_{20}$ -- Gini plane. The lower concentrations of dwarf ETGs, seen in Figure \ref{fig:aVSs_local}, are mirrored in lower values of the Gini coefficient, while dwarf LTGs do not share the low and high values of \textit{M}$_{20}$ and Gini respectively like their massive counterparts. Overall, the \textit{M}$_{20}$ -- Gini plane is unable to cleanly separate the ETG and LTG classes in the dwarf regime like it does in the massive-galaxy regime.

%--------------------------------------------------------------

\subsection{Comparison to massive galaxies in the nearby Universe ($z<0.1$)}

\label{sec:outside}

We proceed by comparing our dwarf sample with massive (\textit{M}$_\star$ > 10$^{10}$ M$_\odot$) galaxies observed in the Dark Energy Survey (DES; \citet{Abbott2018}) at low redshift ($z<0.1$). Note that, given the relatively small ($\sim$2 deg$^2$) area of COSMOS (as opposed to $\sim$5000 deg$^2$ in DES), there are only a handful of massive galaxies in the COSMOS2020 catalogue at $z<0.08$, making it impossible to do a statistical comparison between dwarfs and massive galaxies within this footprint alone. We restrict our DES comparison to massive galaxies only. This is because, while DES is deeper than the SDSS, the magnitude limits of the training set used for the DES classifications ($16<i<18$) mean that DES dwarfs with reliable morphological classifications remain biased towards blue, star-forming objects (in a similar fashion to what is the case for SDSS). As we discuss below, and demonstrate in Appendix \ref{app:DESdwarfs}, this biases the DES dwarfs in the catalogues we use towards blue, morphologically late-type systems. 

Given the upper redshift limit of our DES sample and a median seeing of 0.9 arcseconds for the DES images, the physical scale that is resolved in these images is $\sim$1.6 kpc or better. The DES sample used for our comparisons is constructed by cross-matching three catalogues: 

\begin{itemize}

    \item \citet{Zou2022} provide physical properties of galaxies, including photometric redshifts, stellar masses and SFRs of around 300 million galaxies from DES DR2 data. The photometric redshifts and their uncertainties are calculated using $ugrizyW1W2$ photometry, from the SCUSS \citep{Zou2016}, SDSS \citep{Strauss2002} and WISE \citep{Wright2010} surveys in a hybrid way via local regression on a spectroscopic training set combined with template fitting. The accuracy of the photometric redshifts for galaxies in our mass range of interest is $\sim$0.024 at $z<1.2$. 
    
    \item \citet{Tarsitano2018} provides measurements of morphological parameters including concentration, asymmetry, clumpiness, \textit{M}$_{20}$, Gini coefficient and the S\'ersic index in the $g$, $r$ and $i$ bands for 45 million galaxies from DES Year 1 data. 
    
    \item \citet{Cheng2021} provides one of the largest catalogues of morphological classifications to date, comprising measurements for over 20 million galaxies using the DES Y3 data, based on convolutional neural networks. The training sets used in this work are taken from the Galaxy Zoo 1 (GZ1) catalogue \citep{Lintott2011} which consist of bright galaxies with $16 < i < 18$ at $z<0.25$. The accuracy of the `superior confidence' ($i<18$) classifications, which we use here, reaches 99 per cent for galaxies when compared to the GZ1 data. This makes this catalogue ideal for comparing the morphological properties of our dwarfs to those of massive galaxies in the nearby Universe. The final catalogue provides probabilities for two galaxy types: ETGs and LTGs. 
    
\end{itemize}

For the comparison with our HSC dwarf population, {\color{black} we first cross-match and combine the three catalogues mentioned above. We then use this combined catalogue to select galaxies} with M$_\star$ > 10$^{10}$ M$_{\odot}$ at $z<0.1$, which have probabilities of being ETG or LTG higher than 90 per cent, redshift and stellar mass errors less than 0.02 and 0.2 dex respectively and which have the highest (`superior') confidence flag of 4. Importantly, galaxies with the superior confidence flag span the same magnitude range ($16<i<18$) as that of the Galaxy Zoo training set used in \citet{Cheng2021} and will, therefore, have the most reliable morphological classifications. Our final sample contains 809 objects out of which 103 are ETGs and 706 are LTGs. {\color{black} For consistency with the DES studies described above, we calculate the morphological parameters for our HSC sample using pixels within $R\rm_{petro}$, as in \citet{Tarsitano2018}.}

\begin{figure}
\center
\includegraphics[width=\columnwidth]{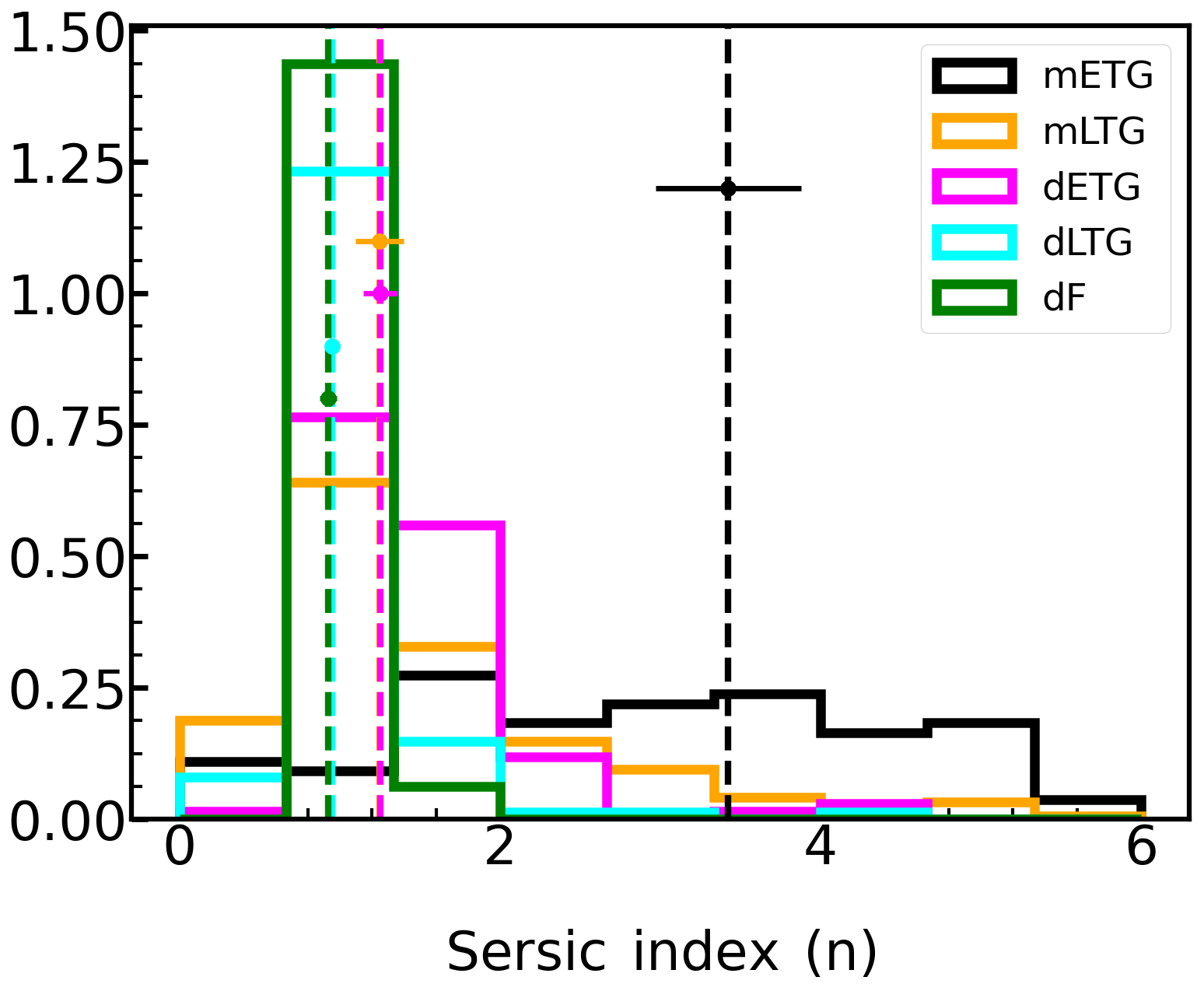}
\caption{S\'ersic index distribution for our dwarf sample and massive galaxies in DES. Median values (together with their associated uncertainties) are shown using the vertical lines and horoizontal error bars. `m' and `d' correspond to the massive (M$_\star$ > 10$^{10}$ M$_\odot$) and dwarf (10$^{8}$ M$_\odot$ < M$_\star$ < 10$^{9.5}$ M$_\odot$) regime, respectively. ETG = early-type galaxy, LTG = late-type galaxy, F = featureless galaxy.}
\label{fig:nVSc}
\end{figure}

\begin{figure}
\center
\includegraphics[width=0.9\columnwidth]{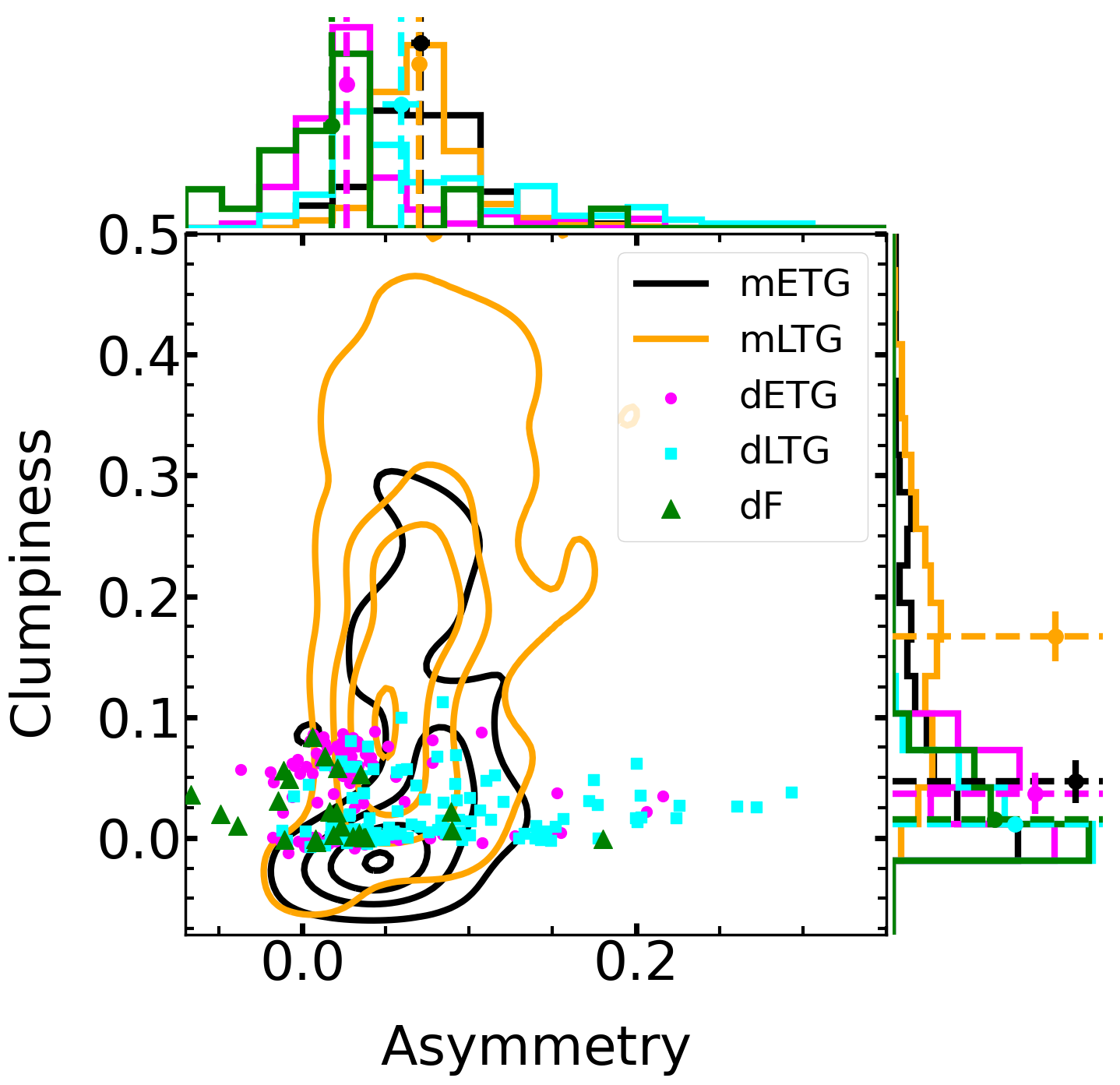}
\includegraphics[width=0.9\columnwidth]{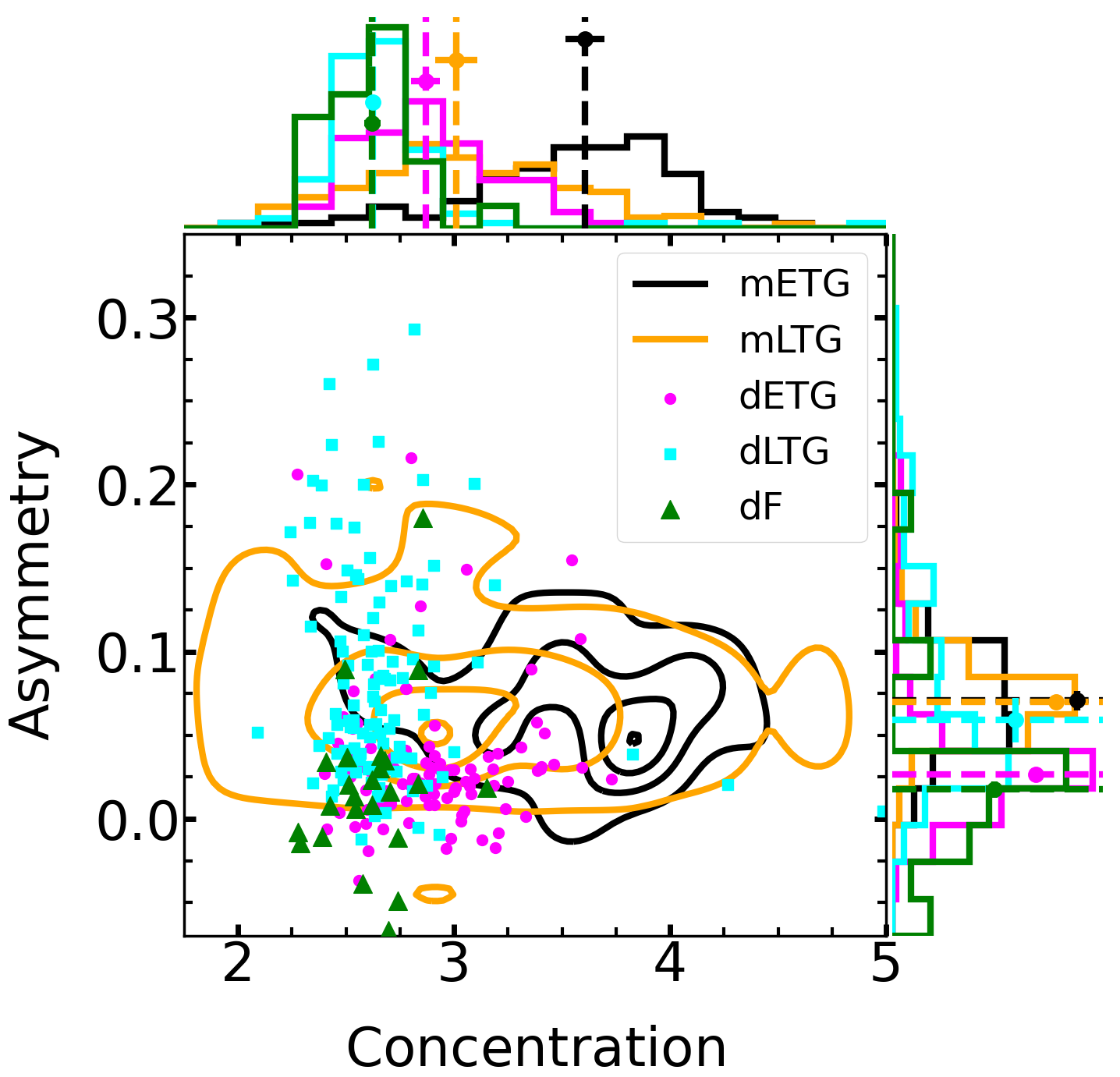}
\includegraphics[width=0.9\columnwidth]{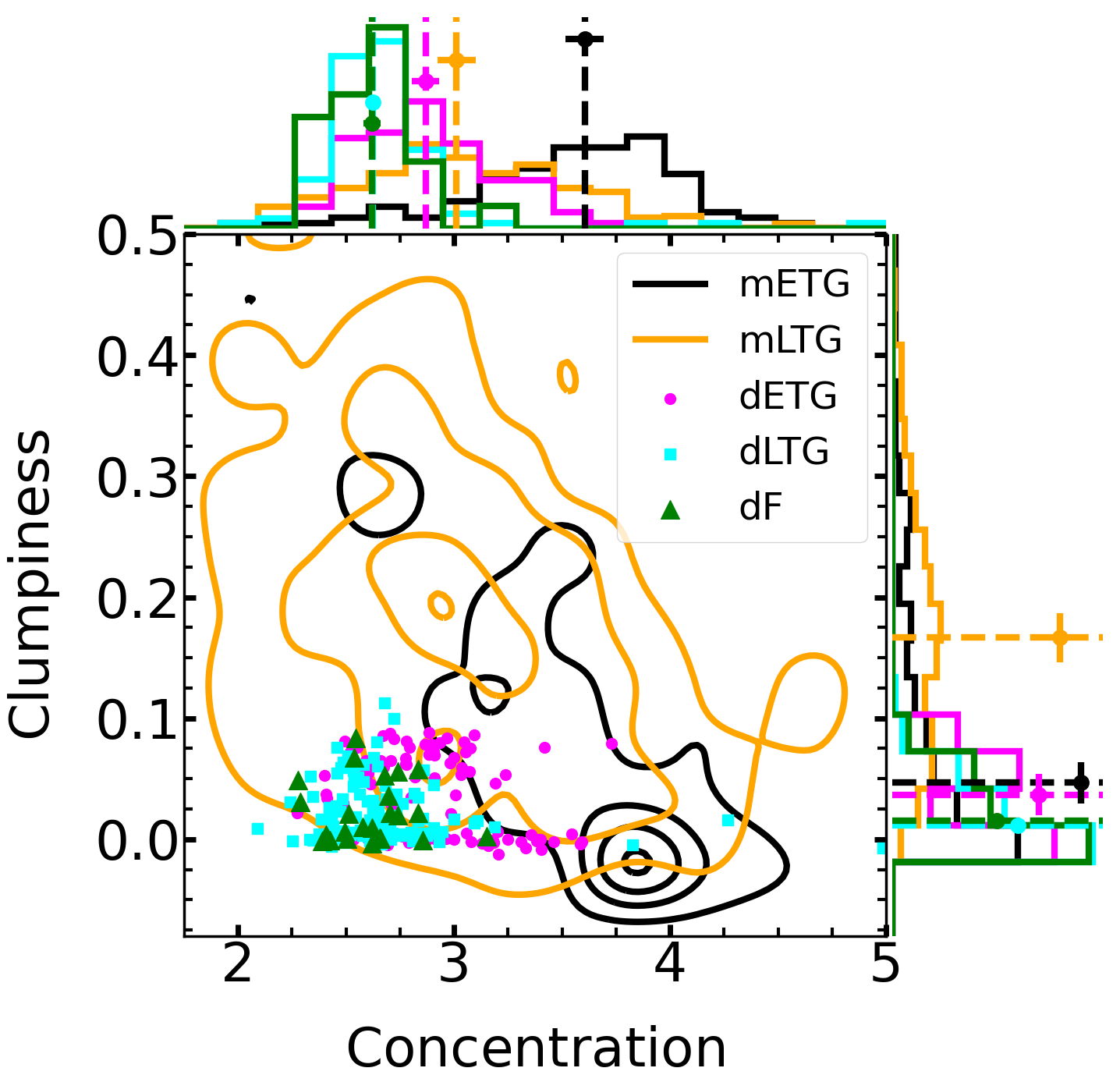}
\caption{Combinations of CAS parameters for various morphological classes in our dwarf sample and massive galaxies in the DES sample (contours). Distributions of parameters and median values (together with their associated uncertainties) are shown on the sides of all panels. `m' and `d' correspond to the massive (M$_\star$ > 10$^{10}$ M$_\odot$) and dwarf (10$^{8}$ M$_\odot$ < M$_\star$ < 10$^{9.5}$ M$_\odot$) regimes, respectively. ETG = early-type galaxy, LTG = late-type galaxy, F = featureless galaxy. Each DES distribution contains 4 equidistant contours starting with 2 counts pixel$^{-1}$ and ending with the maximum pixel value.}
\label{fig:sVSa}
\end{figure}

We begin, in Figure \ref{fig:nVSc}, by presenting the S\'ersic indices of different morphological classes in both the dwarf and massive galaxy regimes. Median values are shown using the dashed vertical lines. The bootstrapped standard errors in the medians are shown using the small horizontal bars. The massive-galaxy regime exhibits the known trends in S\'ersic index, with the ETG population (which has a median of $\sim$3.5) exhibiting larger values than their LTG counterparts (which has a median of $\sim$1.3). {\color{black} As originally noted by \citet{faber83},} unlike their massive counterparts, dwarf ETGs exhibit relatively low values of S\'ersic index (with a median of $\sim$ 1.5 in their study), mirroring the differences in concentration between dwarf and massive ETGs seen earlier in Section \ref{sec:local}). The different dwarf morphological classes span similar values of S\'ersic index, indicating that this parameter is not a reliable discriminator between morphological classes in the dwarf regime.

\begin{figure}
\center
\includegraphics[width=\columnwidth]{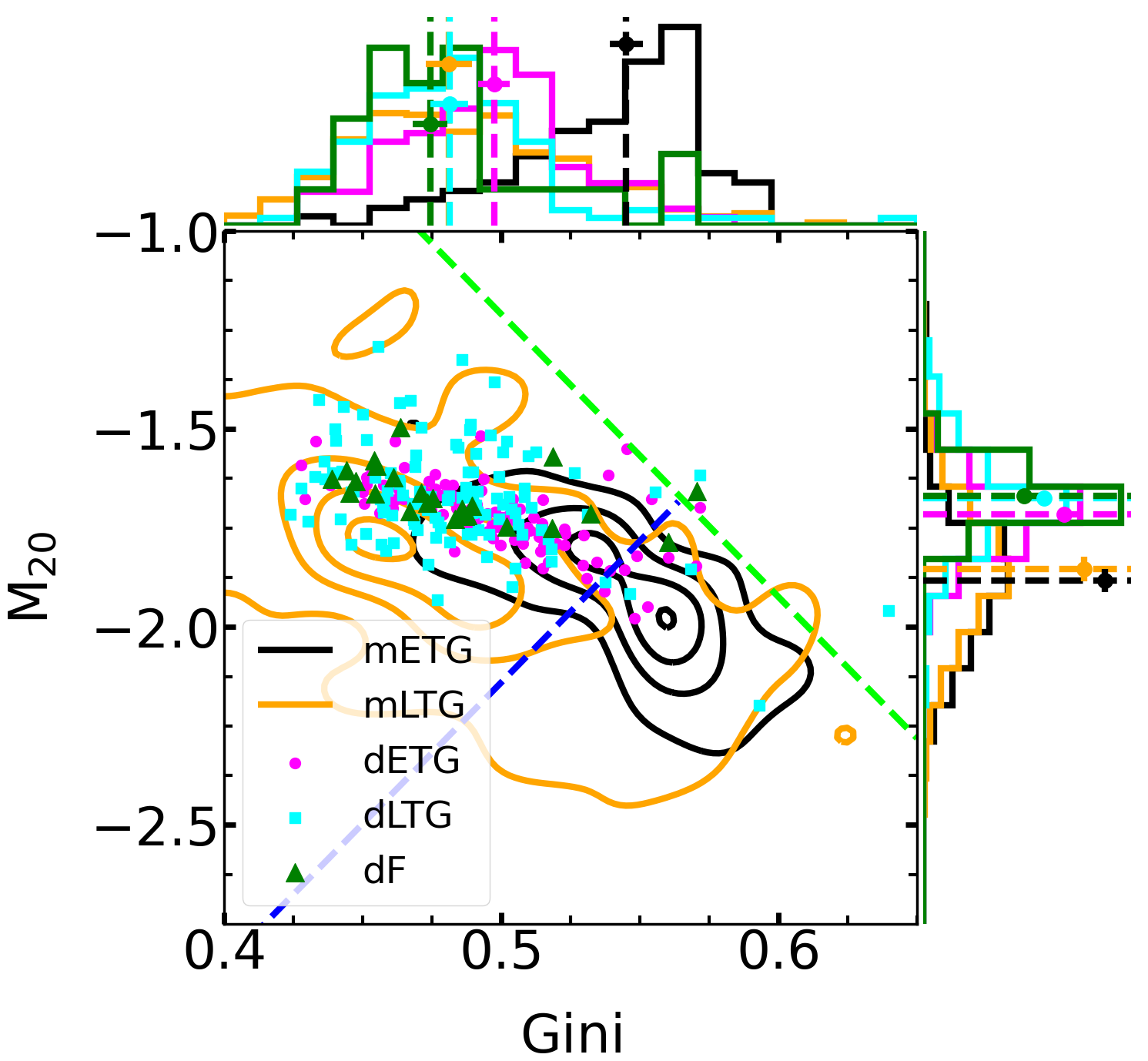}
\caption{\textit{M}$_{20}$ vs Gini for our dwarf sample and massive galaxies in the DES sample (contours). The light green dashed line represents the interaction criterion from \citet{Lotz2008}. The dashed blue line represents the separation relation between ETGs and LTGs described in \citet{Lotz2008}. Distributions of parameters and median values (together with their associated uncertainties) are shown on the sides of all panels. `m' and `d' correspond to the massive (\textit{M}$_\star$ > 10$^{10}$ M$_\odot$) and dwarf (10$^{8}$ M$_\odot$ < \textit{M}$_\star$ < 10$^{9.5}$ M$_\odot$) regimes, respectively. ETG = early-type galaxy, LTG = late-type galaxy, F = featureless galaxy. Each DES distribution contains 4 equidistant contours starting with 2 counts pixel$^{-1}$ and ending with the maximum pixel value.}
\label{fig:m20VSg}
\end{figure}

In Figures \ref{fig:sVSa} and \ref{fig:m20VSg} we compare the morphological parameters of our dwarf sample to the massive galaxy population from DES. The trends we find here are broadly similar to those from the local Universe comparison in Section \ref{sec:local}. But the much larger number of DES galaxies enables us to put these trends on a firmer statistical footing. The top panel of Figure \ref{fig:sVSa} indicates that dwarf ETGs exhibit similar values of asymmetry as their massive counterparts (similar to our findings for the very local Universe in Section \ref{sec:local}). The locations of dwarf LTGs are virtually orthogonal to their massive counterparts in the clumpiness vs asymmetry plane. While dwarf LTGs exhibit relatively high asymmetry and low clumpiness values, massive LTGs show high clumpiness and low asymmetry values. As we noted above, the difference in clumpiness between the dwarf and high mass LTGs, which is seen both in the very local Universe and DES comparisons, could be caused by the higher SFRs in the massive LTGs. Indeed, the median SFR of massive LTGs in DES is a factor of 3 higher than the median SFR of the HSC dwarf LTGs (SFR $\rm_{HSC,dLTG}\sim$0.05 M$_\odot$ yr$^{-1}$; SFR $\rm_{DES,mLTG}\sim$0.15 M$_\odot$ yr$^{-1}$). The fact that clumpiness typically correlates strongly with the SFR (as also seen in e.g. C03) suggests that massive LTGs are likely to show higher values of clumpiness. 

The middle panel of Figure \ref{fig:sVSa} shows that, like in the very local Universe comparison, dwarf ETGs show much lower values of concentration than their massive counterparts, with the median concentration of dwarf ETGs being around a factor of 1.4 smaller than that in the massive ETG population. As noted in the previous section, past studies have shown that the \textit{M}$_{20}$ -- Gini plane can be used to separate ETGs and LTGs in the massive-galaxy regime, using various instruments at different redshifts e.g. HST at intermediate and high redshift \citep[e.g.][]{Lotz2008,Lee2013}, DECaM via DES at $z<0.25$ \citep{Cheng2021} and Spitzer via the S$_4$G survey \citep{Holwerda2014}) at distances up to $\sim$40 Mpc. Figure \ref{fig:m20VSg} shows the ETG -- LTG separation criterion (dashed blue line), defined by \citet{Lotz2008}, applied to the DES galaxies from \citet{Cheng2021} and confirms that this plane can separate massive ETGs and LTGs well. However, dwarf ETGs and LTGs show a reasonable degree of overlap in the \textit{M}$_{20}$ -- Gini plane, which makes it difficult to cleanly separate these morphological classes in the dwarf regime using these parameters.

%--------------------------------------------------------------

\subsection{Can interacting and non-interacting dwarfs be separated using morphological parameters?}
\label{sec:inter2}

We now explore whether morphological parameters could be used to separate interacting dwarfs from their non-interacting counterparts. {\color{black}In the massive-galaxy regime, an asymmetry threshold of $A = 0.35$ has often been used to identify interacting systems during a significant fraction of the interaction \citep[e.g.][]{Conselice2000b,Conselice2003}}. While galaxies such as ULIRGs, which are undergoing significant merging, span a wide range of asymmetry values, non-interacting galaxies in the massive regime appear not to extend beyond $A \sim 0.35$, which makes this a reasonable criterion for selecting strongly interacting systems. 

\begin{figure}
\center
\includegraphics[width=0.9\columnwidth]{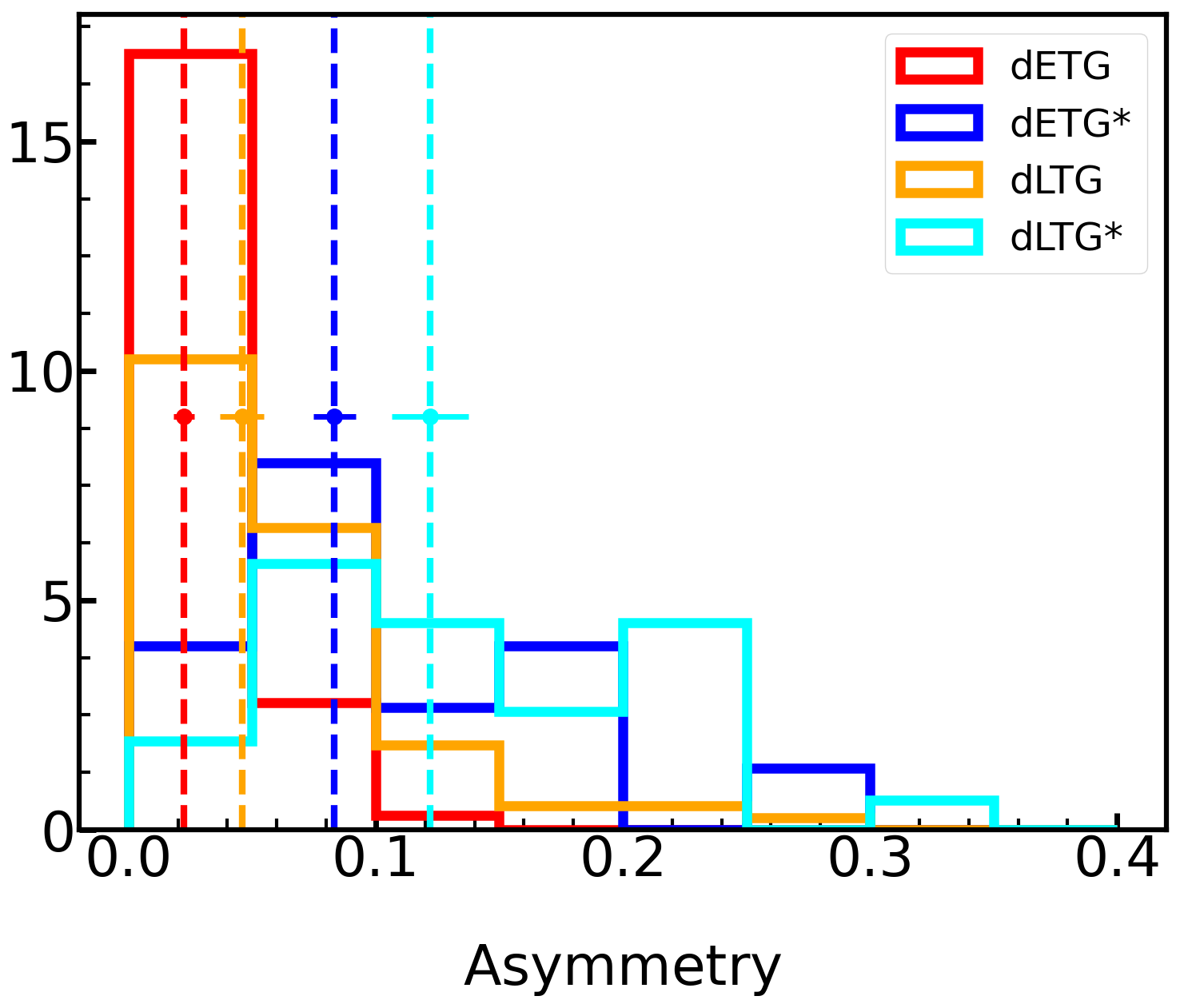}
\includegraphics[width=0.9\columnwidth]{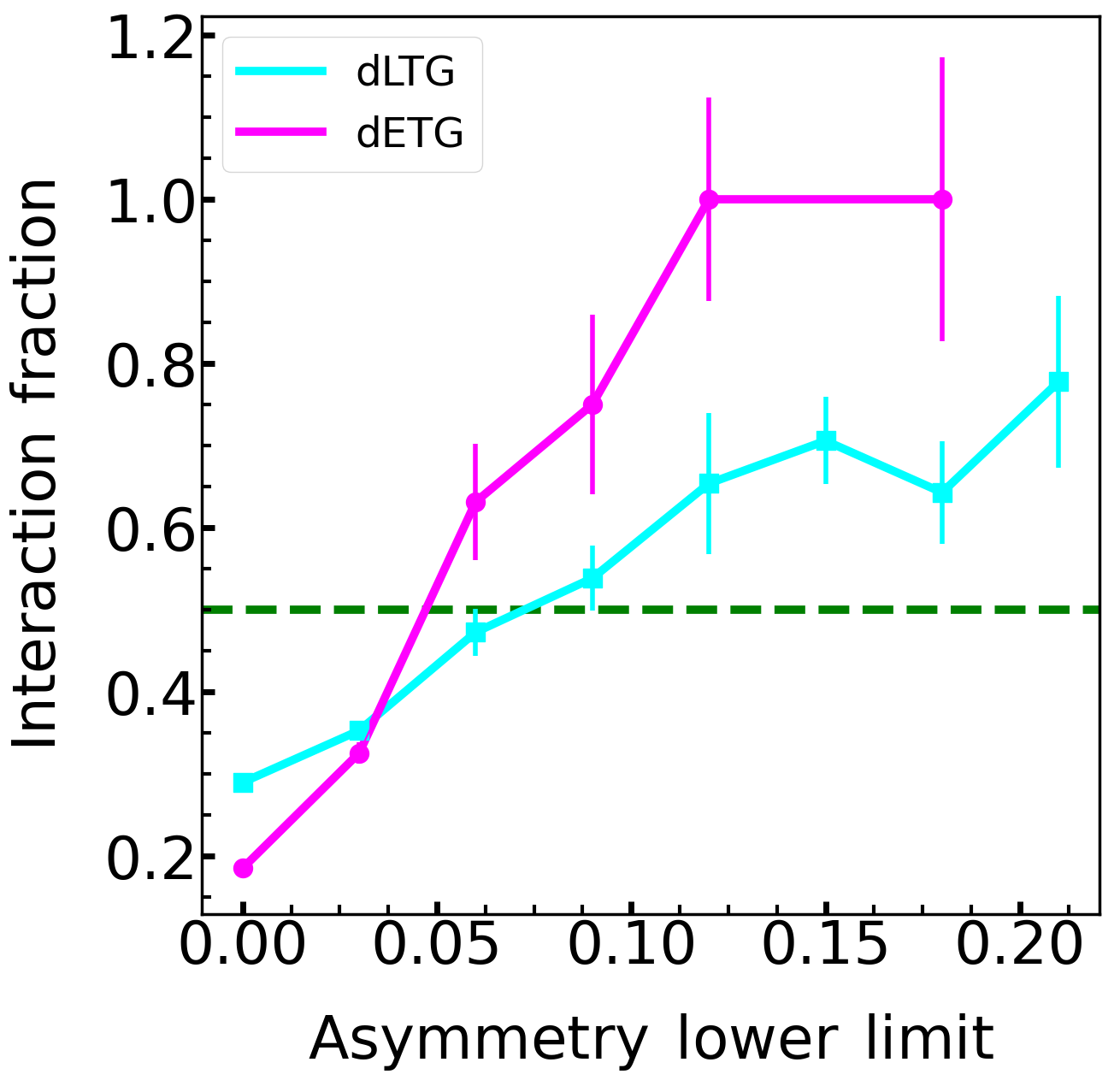}
\caption{\textbf{Top:} Asymmetry distributions for interacting (starred in legend) and non-interacting (not starred) early (dETG) and late (dLTG) type dwarf galaxies. Median values (together with their associated uncertainties) are shown using vertical lines and horizontal error bars. \textbf{Bottom:} Fractions of interacting galaxies for different lower asymmetry thresholds, shown on the x-axis. The green horizontal dashed line indicates an interaction fraction of 50 per cent.}
\label{fig:aVSc_inter}
\end{figure}

In Figure \ref{fig:aVSc_inter}, we compare the values of asymmetry for dwarf galaxies which have been flagged as being interacting to those that have not. {\color{black} The dwarf ETGs and LTGs which are interacting exhibit median asymmetry values that are larger by around a factor of 4 and 3, respectively, than that of their non-interacting counterparts (see the top panel of this figure).} As might be expected, the presence of visually-identifiable morphological disturbances correlates with larger values of asymmetry. There is some overlap between the interacting and non-interacting populations (at least in the relatively small dwarf sample in this study), which means that using asymmetry to cleanly separate disturbed and undisturbed dwarfs is challenging. 

Nevertheless, the fraction of interacting galaxies increases for progressively higher thresholds of asymmetry (see the bottom panel of this figure). {\color{black} For example, in dwarf ETGs which have asymmetry values larger than $\sim$0.05, more than 50 per cent of galaxies are interacting. The corresponding value for dwarf LTGs is $\sim$0.08. These thresholds can be used to select samples of dwarfs in which a majority are likely to be interacting (in images that have similar depth and resolution to the ones used in this study). We note that these thresholds may change for systems outside the stellar mass range probed in this paper (10$^{8}$ M$_{\odot}$ < \textit{M}$_{\rm{\star}}$ < 10$^{9.5}$ M$_{\odot}$) or at higher redshift.}

\begin{figure}
\center
\includegraphics[width=0.9\columnwidth]{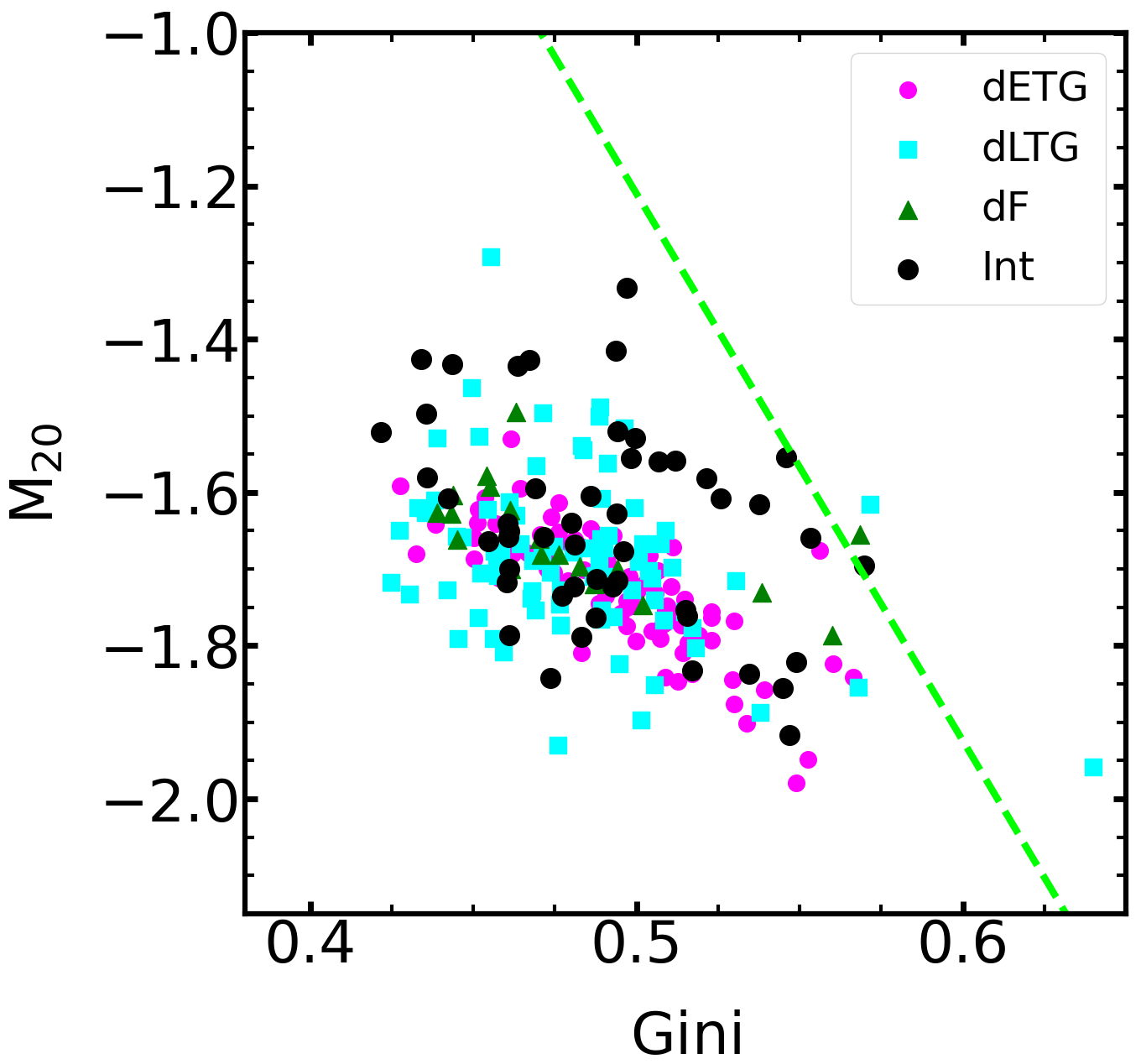}
\caption{\textit{M}$_{20}$ vs Gini coefficient for our dwarf sample. The light green dashed line represents the criterion from \citet{Lotz2008} which is often used to separate interacting galaxies in the massive-galaxy regime. The black circles represent dwarf galaxies which are interacting.}
\label{fig:GvsM20_inter}
\end{figure}

In a similar vein, L04 and \citet{Lotz2008} have formulated an interaction criterion in the \textit{M}$_{20}$ -- Gini plane for massive galaxies, which is represented by the green dashed line in Figure \ref{fig:GvsM20_inter}. Using ULIRGs as proxies for strongly interacting galaxies, they have shown that most ULIRGs reside on the right-hand side of this line. However, this discriminator does not separate interacting and non-interacting dwarfs and performs worse than asymmetry.

%--------------------------------------------------------------

\section{Implications for the evolutionary histories of dwarf morphological classes}

\label{sec:implications}

In this section, we complete our study by bringing together our results and discussing their implications for the evolutionary histories of the various dwarf morphological classes. 

\subsection{Dwarf ETGs have different evolutionary histories to massive ETGs}

\label{sec:dwarf_ETG_formation}

{\color{black}Section \ref{sec:inter} indicates several interesting differences between dwarf and massive ETGs. Contrary to what is found in the massive-galaxy regime, the incidence of interactions in dwarf ETGs is lower than that in dwarf LTGs. The interaction fraction in dwarf ETGs appears to be around a factor of 5 lower than that in their massive counterparts. In a similar vein, the frequency of dust lanes, which are signposts of interactions with lower mass companions, are several factors lower in dwarf ETGs than in massive ETGs.} While the majority of massive ETGs have redder colours than their LTG counterparts, the majority of dwarf ETGs are blue like their LTG counterparts (Section \ref{sec:colours}), due to a high incidence of features such as blue cores. Finally, Section \ref{sec:params} demonstrates that dwarf ETGs have lower concentrations, S\'ersic indices and Gini coefficients (all of which are measures of how concentrated the light profile is within a galaxy) than massive ETGs. The strong differences in both the structural properties and rest-frame colours (which trace recent star formation histories) suggest that the evolutionary histories of dwarf ETGs is likely to be different from that of their massive counterparts. 

The lower incidence of interactions and the less concentrated light profiles both suggest that the origin of dwarf ETGs might have less to do with interactions and merging than their massive counterparts. This appears to be in line with the findings of recent observational work that suggests that secular accretion from the cosmic web is the dominant evolutionary channel for these systems \citep{Lazar2023}. This observational picture also appears consistent with theoretical studies which have explored the formation mechanisms of dwarf ETGs. 

For example, some studies postulate the formation of dwarf ETGs through monolithic collapse, with the subsequent star formation history regulated by internal processes like stellar and supernova feedback \citep[e..g][]{Yoshii1987,Chiosi2002}. Other studies postulate a `harassment' scenario, in which dwarf ETGs are created from dwarf LTGs as they orbit massive galaxies. This takes place both via tidal perturbations, which remove angular momentum and make the system more dispersion dominated, and the tidal stripping of gas \citep[e.g.][]{Mayer2001,Marcolini2003}. While such a scenario is more likely in dense environments like groups and clusters, it is plausible that it also operates is relatively less dense environments like the ones explored in this study. A further potential channel of dwarf ETG formation is in the tidal tails created by the gas-rich interactions of massive galaxies \citep[e.g.][]{Duc2004}, although the overall fraction of dwarfs that is likely to be created via this channel is expected to be very small \citep[e.g.][]{Kaviraj2012} and a tidal origin does not easily explain the formation of large numbers of dwarf ETGs in low-density environments (such as the ones studied here).

%--------------------------------------------------------------

\subsection{Featureless dwarfs in low-density environments are formed via baryonic feedback}

\label{sec:fdwarfs}

Section \ref{sec:visual} indicates that a minority ($\sim$10 per cent) of our dwarfs fall in the featureless category. Interestingly, all of these systems are in the lower half of the dwarf stellar mass range considered in this study (10$^8$ M$_{\odot}$ < \textit{M}$_\star$ < 10$^{8.75}$ M$_{\odot}$), suggesting that the formation of these featureless galaxies becomes progressively easier as stellar mass decreases. This, in turn, suggests that the formation of featureless galaxies is driven by processes whose impact becomes stronger as the depth of the gravitational potential well becomes shallower e.g. environmental processes like tidal perturbations and ram pressure or baryonic feedback from supernovae or AGN.  

However, the featureless systems are found at similar local densities as the other morphological classes (Section \ref{sec:env}) and we have shown that the COSMOS2020 footprint does not contain dense environments in the redshift range of our study (Section \ref{sec:sample}). 
Finally, Section \ref{sec:colours} shows that the featureless galaxies exhibit a range of rest-frame colours, indicating that they are not quenched as a population and that many show evidence for recent star formation. These points suggest that featureless galaxies in low-density environments, at least in our dwarf sample, are likely to be created, not by environmental processes but by internal mechanisms such as baryonic feedback, either from supernovae and/or AGN \citep[e.g.][]{Kaviraj2019,Koudmani2022}.

%--------------------------------------------------------------

\subsection{Shallower potential wells cause late-type dwarfs to be structurally more asymmetric than massive late-types}

Figure \ref{fig:sVSa} indicates that dwarf LTGs show higher median asymmetry than their massive counterparts. They also extend to much larger asymmetry values than massive LTGs. This is likely caused by the fact that the shallower potential wells in the dwarf LTGs make them more susceptible to internal displacement of material. For example, baryonic feedback may be able to move gas around more easily, making the gas reservoir more asymmetric. Stars that form from this gas will then inherit this asymmetry. Alternatively, tidal perturbations due to nearby large-scale structure are likely to be able to alter the stellar distribution in dwarfs more easily, also inducing larger asymmetries.  

%--------------------------------------------------------------

\section{Summary}
\label{sec:summary}

Dwarf galaxies dominate the galaxy number density at all epochs, making them fundamental to understanding the evolution of the Universe. While dwarfs have been studied in detail in the very local Universe (e.g. within around 50 Mpc), typical dwarfs are difficult to study at cosmological distances in past large surveys (e.g. the SDSS), because they are too shallow. Outside the local neighbourhood, shallow surveys typically detect dwarfs that have high SFRs, which temporarily boost the luminosity of these galaxies, lifting them above the detection threshold of shallow surveys. However, this also makes the small subset of dwarfs that exist in such surveys biased which, in terms of morphology, skews these samples towards predominantly blue late-type galaxies (as we have shown in Appendix \ref{app:DESdwarfs}).

Quantifying the morphological mix of dwarfs, outside our immediate neighbourhood, in low-density environments requires surveys that are both deep and wide and which can therefore provide unbiased statistical samples of dwarf galaxies. Here, we have constructed such a sample of 257 dwarfs, which lie in the stellar mass range 10$^{8}$ M$_{\odot}$ < \textit{M}$_{\rm{\star}}$ < 10$^{9.5}$ M$_{\odot}$ and in the redshift range $z<0.08$. We have first performed visual inspection, using ultra-deep HSC images (and their unsharp-masked counterparts) that are around 5 magnitudes deeper than standard-depth SDSS imaging, to establish the principal morphological classes in the dwarf regime. We have then explored the local densities, the role of interactions, rest-frame colours and the incidence of bars in these morphological classes. This is followed by an exploration of commonly-used morphological parameters in the dwarf regime and a comparison to both dwarf and massive galaxies in the very local Universe (within 50 Mpc) and the massive-galaxy population in the nearby Universe, using DES at $z<0.1$. The overall aim has been to provide a pilot study that offers a useful benchmark for the study of dwarf morphology using new and forthcoming deep-wide surveys such as Euclid and LSST. Our main results are as follows:

\begin{itemize}
    
    \item Visual inspection reveals three broad morphological classes in the dwarf population in our stellar mass range of interest (10$^{8}$ M$_{\odot}$ < \textit{M}$_{\rm{\star}}$ < 10$^{9.5}$ M$_{\odot}$). Around 43 and 45 per cent of our dwarfs exhibit early-type (elliptical/S0) and late-type (spiral) morphology respectively. Around 10 per cent of dwarfs populate a `featureless' class, that lacks both the central light concentration seen in early-types and any spiral structure. While the dwarf ETGs and LTGs are visually similar to the ETG and LTG classes found in massive galaxies, the featureless class does not have a counterpart in the massive-galaxy regime.   
    
    \item {\color{black}Dwarf ETGs diverge strongly from their massive counterparts in both their structural and photometric properties. They show an incidence of interactions and dust lanes which are several factors lower than that seen in massive ETGs. They are also significantly less concentrated and, unlike massive ETGs, share similar rest-frame colours as dwarf LTGs. This suggests that, unlike their massive counterparts, the formation of dwarf ETGs may be driven less by interactions and more by secular processes over cosmic time.} 

    \item The COSMOS footprint does not contain large groups or clusters at $z<0.08$ and the local density of the various dwarf morphological classes, traced via projected distances to the first to the tenth nearest massive neighbours, do not show significant differences. This implies that the creation of the featureless dwarfs in low-density environments is likely to be driven by internal baryonic (stellar or AGN) feedback rather than by environmental processes.

    \item Around 20 per cent of the star formation activity in dwarfs, in the stellar mass and redshift range probed in this study, is likely to be driven by interactions. 

    \item The (strong) bar fraction in dwarf galaxies is around 11 per cent, consistent with the recent literature and lower than that found in the massive-galaxy regime ($\sim$20 per cent). Unlike the massive galaxy regime, the median rest-frame colour of (strongly) barred dwarfs is not significantly redder than that of unbarred dwarfs. 
    
    \item It is challenging to separate different dwarf morphological classes using commonly-used parameters like the S\'ersic index, CAS, \textit{M}$_{20}$ and the Gini coefficient. This is largely driven by the fact that dwarf ETGs are less concentrated than their massive counterparts. Thus, the light concentration, which is a key separator between morphological classes in the massive-galaxy regime, does not offer the same level of discrimination in the dwarf regime. 
    
    \item {\color{black} The asymmetry of dwarf ETGs and LTGs that are interacting is larger, by approximately factors of 4 and 3 respectively, than their non-interacting counterparts. Asymmetry thresholds of 0.05 and 0.08 respectively are able to select samples of dwarf ETGs and LTGs in which more than 50 per cent of the galaxies are interacting (in images that have similar resolution and depth to the ones used in this study).} 
    
\end{itemize}

Broadly speaking, there is evidence that a transition in morphological behaviour occurs between the dwarf and massive regimes (i.e. around \textit{M}$_{\rm{\star}}$ $\sim$ 10$^{9.5}$ M$_{\odot}$). While similar morphological classes (ETGs and LTGs) exist in both regimes, they have different formation mechanisms and star formation histories. In addition, new morphological classes, such as the featureless objects emerge in the dwarf regime, which do not share the characteristics of the classical ETG and LTG populations. It is likely that galaxies may show even more morphological diversity at stellar masses lower than the limit of our study (i.e. at \textit{M}$_{\rm{\star}}$ < 10$^{8}$ M$_{\odot}$). 

Our study poses several questions that need to be addressed in future work. What is the relationship between the different dwarf morphological classes? Can dwarf LTGs transform into dwarf ETGs as is seen in the massive-galaxy regime and, if so, via what processes? Could the featureless dwarfs be short-lived systems, as might be suggested by their low number fractions? Or do they transform, over short timescales, into the other morphological types? How do these relationships vary as a function of stellar mass and environment? Answering these questions requires statistical explorations of how the morphological mix of dwarfs evolves over cosmic time. While this is beyond the scope of this study, it will form the basis of future investigations using both ground and space-based data from surveys like LSST and Euclid.

Given the time-consuming nature of visual inspection, the challenges in separating dwarf morphological classes using parameters and the fact that millions of dwarfs will be imaged by forthcoming deep-wide surveys (e.g. LSST and Euclid), future statistical studies of dwarf morphology will likely need to leverage machine-learning techniques. Given the differences between the dwarf and massive regimes, there is a need to construct training sets (e.g. using systems like Galaxy Zoo) that can be used for dwarf classifications based on supervised machine-learning \citep[e.g.][]{Walmsley2022}. Unsupervised machine learning techniques \citep[e.g.][]{Martin2020,Lazar2023}, which can perform morphological clustering of arbitrarily large galaxy populations from wide-area surveys, are also likely to be important for future morphological studies in the dwarf regime. 

To conclude, studying the evolution of dwarfs over the lifetime of the Universe, using statistically significant samples of dwarf galaxies from deep-wide surveys like LSST and Euclid, is likely to become a significant endeavour in the coming decades. Such studies will be capable of bringing revolutionary advances in our understanding of the physical processes that drive galaxy evolution. 

%--------------------------------------------------------------

\section*{Acknowledgements}
We are grateful to the referee, Benne Holwerda, for many insightful comments that improved the quality of the paper. We thank Pierre-Alain Duc, Elizabeth Sola and Liza Sazonova for many interesting discussions. IL acknowledges a PhD studentship from the Centre for Astrophysics Research at the University of Hertfordshire. SK and AEW acknowledge support from the STFC [grant numbers ST/S00615X/1 and ST/X001318/1]. SK also acknowledges a Senior Research Fellowship from Worcester College Oxford. 
%--------------------------------------------------------------

\section*{Data Availability}
The morphological parameters produced in this work are available via common online repositories. They can also be obtained by contacting the authors.

%--------------------------------------------------------------

\bibliographystyle{mnras}
\bibliography{paper}

%%%%%%%%%%%%%%%%%%%%%%%%%%%%%%%%%%%%%%%%%%%%%%%%%%

%%%%%%%%%%%%%%%%% APPENDICES %%%%%%%%%%%%%%%%%%%%%

\appendix

\section{Dwarfs in the DES comparison sample are biased towards blue, late-type galaxies}
\label{app:DESdwarfs}

\begin{figure}
\center
\includegraphics[width=0.85\columnwidth]{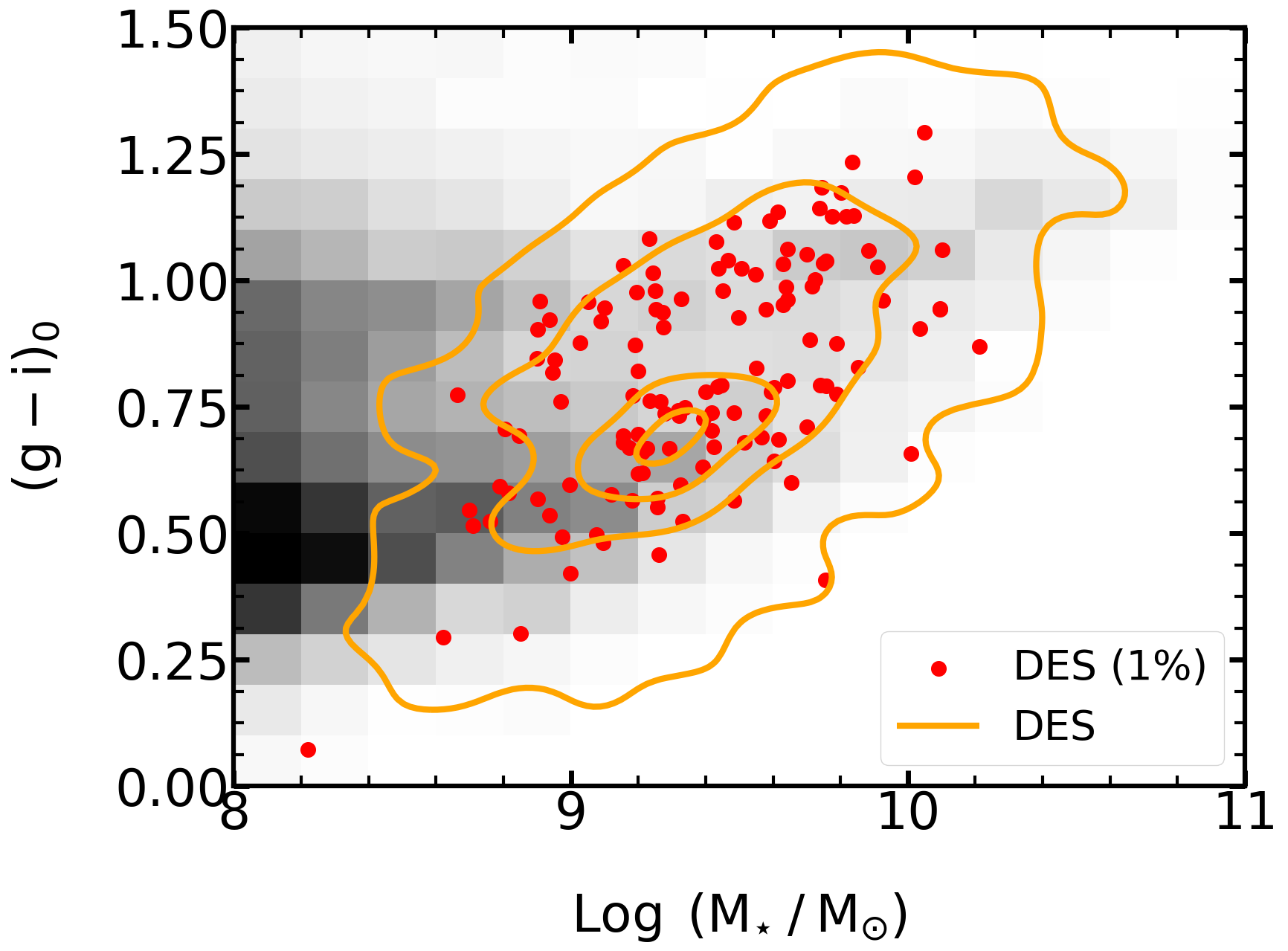}
\caption{$(g-i)$ colour vs stellar mass for galaxies in the COSMOS2020 catalogue (shown using the heatmap) compared to the DES galaxies (shown using the orange contours) described in Section \ref{sec:outside} (i.e. those which have probabilities of being ETG or LTG higher than 90 per cent, redshift and stellar mass errors less than 0.02 and 0.2 dex respectively and which have the highest morphological confidence flag of 4). A random 1 per cent of the DES galaxies are shown overplotted using red points.}
\label{fig:DESvsHSC}
\end{figure}

As noted in Section \ref{sec:outside}, DES dwarfs that conform to the selection criteria we use to construct the DES comparison sample are strongly biased towards, blue galaxies which are dominated by late-type morphologies. This prevents us from comparing our dwarfs to those in DES (as a result of which our DES comparison is restricted to massive galaxies only). Here, we demonstrate these biases. Figure \ref{fig:DESvsHSC} presents the rest-frame $(g-i)$ colour vs stellar mass for HSC galaxies in the COSMOS2020 catalogue (shown using the heatmap) compared to the DES galaxies which satisfy the selection criteria described in Section \ref{sec:outside} i.e. which have probabilities of being ETG or LTG higher than 90 per cent, redshift and stellar mass errors less than 0.02 and 0.2 dex respectively and the highest confidence flag for morphological classifications of 4. The DES population is shown using the orange contours. A random 1 per cent of the DES galaxies are shown overplotted using red points. While massive galaxies are visible in DES regardless of whether they are red or blue, this is not the case for dwarf galaxies. The DES galaxies become progressively bluer at lower stellar masses. Indeed, towards the lower end of the stellar mass range considered in this study, the small fraction of DES objects that do appear in the DES comparison sample are some of the most extreme in terms of their blue colour. 

\begin{figure}
\center
\includegraphics[width=0.8\columnwidth]{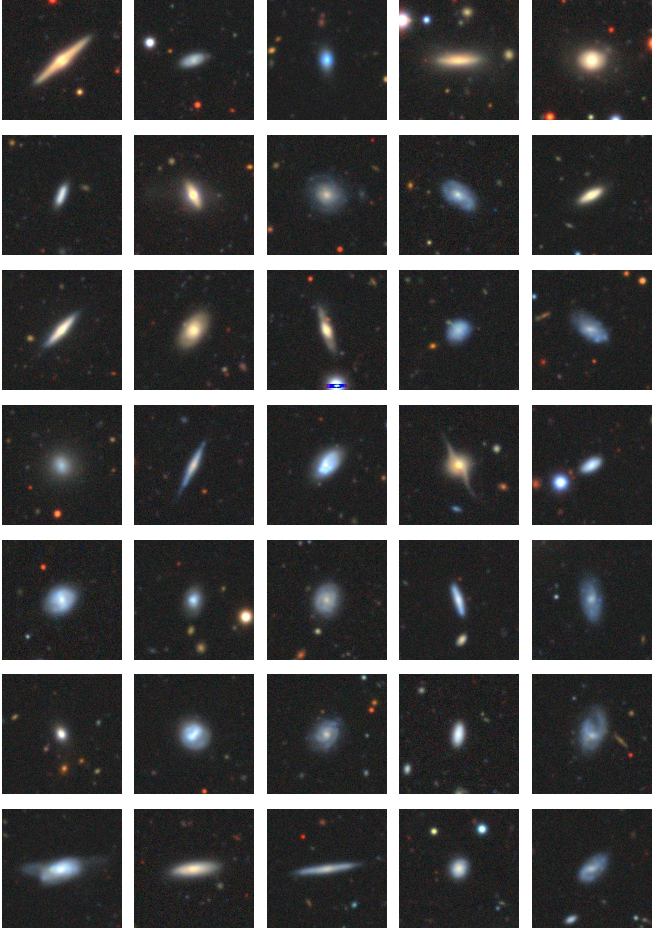}
\caption{$gri$ colour composite images of a random sample of DES dwarfs which satisfy the selection criteria for the DES comparison sample i.e. galaxies which have probabilities of being ETG or LTG higher than 90 per cent, redshift and stellar mass errors less than 0.02 and 0.2 dex respectively and the highest confidence flag for morphological classifications of 4.} 
\label{fig:DESdwarfs}
\end{figure}

Figure \ref{fig:DESdwarfs} shows a random sample of 35 DES dwarfs. Visual inspection of a random sample of 300 galaxies in this DES dwarf population (using $gri$ colour composite images) indicates that the ETG fraction is $\sim$11 per cent, a factor of 4 lower than the ETG fraction observed in the significantly deeper COSMOS2020 catalogue ($\sim$44 per cent, see Section \ref{sec:visual}). Not unexpectedly, a consequence of the DES dwarfs being dominated by blue galaxies is that the morphological mix is also skewed towards late-type systems. Note that reducing the \texttt{confidence\_flag} parameter to 3 does not change these results.

%%%%%%%%%%%%%%%%%%%%%%%%%%%%%%%%%%%%%%%%%%%%%%%%%%

% Don't change these lines
\bsp	% typesetting comment
\label{lastpage}
\end{document}